\documentclass[aps,twocolumn,floats,prl,longbibliography]{revtex4-2}
\usepackage[colorlinks,linkcolor=black,citecolor=blue,urlcolor=black]{hyperref}
\usepackage{amsmath,amsfonts,amssymb,graphicx,bm,color,subfigure,soul}

\begin{document}

\author{G. Spada$^1$, S. Pilati$^{2,3}$ and S. Giorgini$^1$\vspace*{10pt}}

\affiliation{
    $^1$ Pitaevskii Center on Bose-Einstein Condensation, CNR-INO and Dipartimento di Fisica, Universit\`a di Trento, 38123 Povo, Trento, Italy \\
    $^2$ School of Science and Technology, Physics Division, Universit\`a di Camerino, 62032 Camerino, Italy \\
    $^3$ INFN, Sezione di Perugia, I-06123 Perugia, Italy
}

\title{Attractive Solution of Binary Bose Mixtures:\\Liquid-Vapor Coexistence and Critical Point}

\begin{abstract} We study the thermodynamic behavior of attractive binary Bose mixtures using exact path-integral Monte Carlo methods. Our focus is on the regime of interspecies interactions where the ground state is in a self-bound liquid phase, stabilized by beyond mean-field effects. We calculate the isothermal curves in the pressure vs density plane for different values of the attraction strength and establish the extent of the coexistence region between liquid and vapor using the Maxwell construction. Notably, within the coexistence region, Bose-Einstein condensation occurs in a discontinuous way as the density jumps from the normal gas to the superfluid liquid phase. Furthermore, we determine the critical point where the line of first-order transition ends and investigate the behavior of the density discontinuity in its vicinity. We also point out that the density discontinuity at the transition could be observed in experiments of mixtures in traps.
\end{abstract}
\pacs{05.30.Fk, 03.75.Hh, 03.75.Ss} \maketitle

Ultracold atomic samples of Bose particles with attractive interactions typically exhibit instability and collapse of the gaslike phase~\footnote{See {\it e.g.} Ref.~\cite{book} Sec.~11.6}. However, recent developments in the fields of dipolar systems and quantum mixtures have revealed that attractive interactions can also lead to the formation of an exotic liquidlike phase, in which self-bound droplets maintain their shape even without external confinement. The density in these quantum droplets saturates at values 7 orders of magnitude lower than water or liquid helium, but 1 order of magnitude higher than typical Bose condensates. In the case of dipolar gases, the characteristic attractive anisotropic interaction is balanced by short-range, spherically symmetric repulsive forces leading to the formation of elongated droplets that are stable above a critical number of atoms~\cite{Schmitt2016,PhysRevX.6.041039}. Notably, beyond mean-field effects contributing to repulsive interactions are crucial to explain the stabilizing mechanism of these droplets~\cite{PhysRevX.6.041039}.
A similar mechanism is responsible for the formation of quantum droplets in binary Bose mixtures featuring repulsive intraspecies and attractive interspecies couplings~\cite{Science.359.301,PhysRevLett.120.235301,PhysRevResearch.1.033155}. The physical picture, in this case, involves an overall mean-field attraction balanced by a beyond mean-field repulsion. This combination leads to a minimum of the ground-state energy corresponding to the equilibrium density of the liquid state~\cite{PhysRevLett.115.155302}. This is the central density of the droplet, reached if the number of particles exceeds a critical value and remaining constant if more particles are added to the droplet. The observed droplet size and critical atom number agree reasonably well with the ground-state scenario described in Ref.~\cite{PhysRevLett.115.155302}.

The stabilizing mechanism produced by beyond mean-field effects in quantum droplets at $T=0$ has been also confirmed by more microscopic calculations based on the quantum Monte Carlo method both for dipolar~\cite{PhysRevLett.117.205301,PhysRevResearch.1.033088} and short-range interactions~\cite{PhysRevB.97.140502,PhysRevA.99.023618,PhysRevA.104.033319}, including in this latter case low-dimensional systems~\cite{PhysRevLett.117.100401,PhysRevLett.122.105302,PhysRevA.102.023318}. Additionally, the persistence of droplet states in trapped configurations at finite temperature has been verified through numerical simulations using the path-integral Monte Carlo (PIMC) technique~\cite{doi:10.7566/JPSJ.85.053001}.
It has also been found that liquid droplets in vacuum are unstable against thermal fluctuations \cite{10.21468/SciPostPhys.9.2.020,Wang_2020,PhysRevA.103.043316}.
Furthermore, these works point out the difficulty of developing a finite-temperature theory of this liquid state following the standard Bogoliubov scheme.
Specifically, density wave excitations become complex and cannot be used to build a proper thermodynamic description~\footnote{Notice that one-dimensional mixtures do not suffer from this shortcoming~\cite{PhysRevA.103.043316}}.
Because of the challenges in developing a proper theoretical framework for this exotic liquid state of matter, its intriguing aspects, such as the liquid-vapor coexistence line characterizing the first-order phase transition and the critical point where the line terminates and the transition is continuous, have remained largely unexplored.

In this Letter we study the thermodynamic behavior of binary attractive Bose mixtures using exact PIMC methods and we map out the phase diagram in the temperature-density as well as temperature-pressure plane. We determine the region where gas and liquid states coexist in equilibrium and we characterize the corresponding critical point in terms of critical temperature, pressure, and density~\footnote{We note that liquid-gas coexistence states have been predicted at $T=0$ in mixtures with spin imbalance or coherent coupling in Refs~\cite{PhysRevA.107.L031303} and~\cite{he2022quantum}}\nocite{PhysRevA.107.L031303,he2022quantum}. We find that the qualitative behavior of isothermal curves in the coexistence region of densities shares many analogies with the liquid-gas transition in classical fluids. Remarkably, in our quantum degenerate mixture, this first-order transition links the normal gaseous phase and the superfluid liquid phase across the Bose-Einstein condensation of the two components. Both the discontinuous density jump at the transition and the sudden appearance of a finite condensate density could be observed in experiments with trapped mixtures.

We consider a mixture of $N=N_1+N_2$ particles belonging to two distinguishable Bose components with equal mass $m$ described by the Hamiltonian $H$, where particle coordinates ${\bf r}_i$ and ${\bf r}_{i^\prime}$ refer to the first and second component:
\begin{eqnarray}
	H&=&-\frac{\hbar^2}{2m}\sum_{i=1}^{N_1}\nabla_i^2
	-\frac{\hbar^2}{2m}\sum_{i^\prime=1}^{N_2}\nabla_{i^\prime}^2 + \sum_{i<j} ^{N_1} v(|{\bf r}_i-{\bf
	r}_j|) \nonumber\\ &+& \sum_{i^\prime<j^\prime}^{N_2} v(|{\bf r}_{i^\prime}-{\bf r}_{j^\prime}|) +
	\sum_{i,i^\prime}^{N_1,N_2} v_{12}(|{\bf r}_i-{\bf r}_{i^\prime}|) \;.
	\label{hamiltonian}
\end{eqnarray}
Particles interact via the interspecies potential $v_{12}(r)$ and equal repulsive intraspecies potentials $v(r)$. For the first we use a zero-range pseudopotential featuring a negative $s$-wave scattering length $a_{12}$, while for the latter hard-sphere interactions are implemented: $v(r)=+\infty$ if $r<a$ and zero otherwise. Furthermore, the two populations are balanced: $N_1=N_2=N/2$ and calculations are performed at fixed overall density $n=N/V$ in a cubic box of volume $V$ with periodic boundary conditions.
We use the implementation of the PIMC algorithm detailed in~\cite{condmat7020030}, which is particularly well suited for dealing with systems subject to periodic boundary conditions, further extended to the case of attractive mixtures~\footnote{See Supplemental Material, which includes Refs.~\cite{RevModPhys.67.279,PhysRevA.105.013325,PhysRevLett.96.070601,doi:10.1063/1.463076,Alder1962dis,reif2010history,PhysRevLett.29.28,10.1063/1.4720089,Rovere_1990}}. \nocite{RevModPhys.67.279,PhysRevA.105.013325,PhysRevLett.96.070601,doi:10.1063/1.463076,Alder1962dis,reif2010history,PhysRevLett.29.28,10.1063/1.4720089,Rovere_1990}
%
Calculations are performed at fixed temperature $T$ and for varying densities expressed in terms of the gas parameter $na^3$. The interspecies $s$-wave scattering length $a_{12}$ is chosen to be in the mean-field theory unstable region $|a_{12}|/a>1$, but close to the threshold. In three spatial dimensions, this region features a stable liquid phase and corresponds to realistically small values of the gas parameter~\cite{PhysRevLett.115.155302} while it avoids strong effects arising from the hard-sphere potential used to model the intraspecies interactions and ensures the universality~\cite{PhysRevA.74.043621}.

\begin{figure}
	\centering
	\includegraphics[width=0.85\columnwidth]{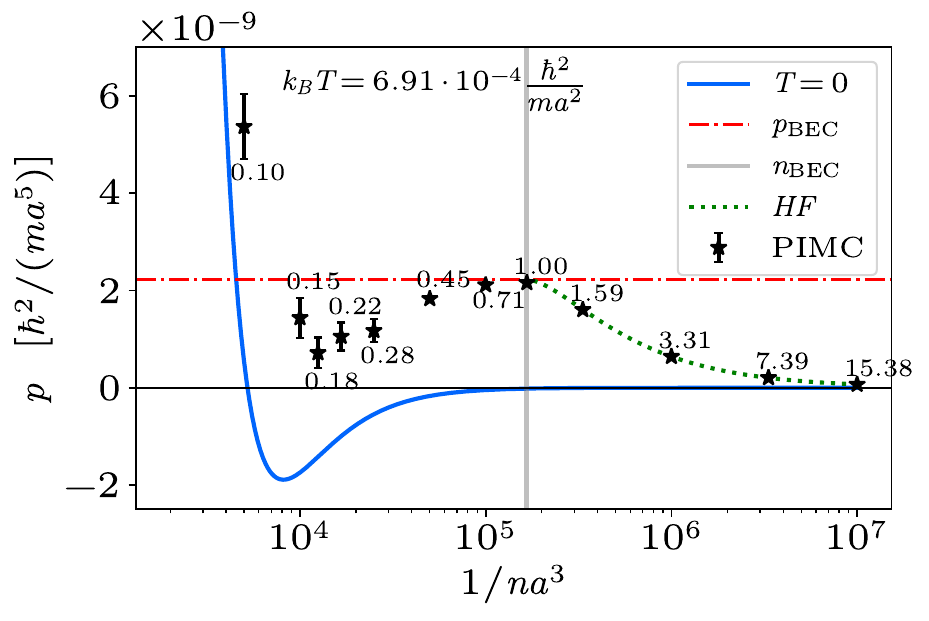}\vspace{-5pt}
	\caption{Isothermal curves of pressure as a function of density. The solid line refers to the $T=0$ result from Eq.~(\ref{Petrov}) while the dotted line is the Hartree-Fock (HF) prediction from Eq.~(\ref{HF}) holding for $n<n_{\mathrm{BEC}}$. The PIMC results correspond to $N=216$ particles. The vertical and horizontal dot-dashed lines refer respectively to the density $n_{\mathrm{BEC}}$ and pressure $p_{\mathrm{BEC}}$ at the onset of BEC. Close to each PIMC point we report the corresponding value of the reduced temperature $T/T_{\mathrm{BEC}}$, where $n\lambda_{T_{\mathrm{BEC}}}^3=2\zeta(3/2)$. }
	\label{fig1}
\end{figure}

In Fig.~\ref{fig1} the case $a_{12}/a=-1.2$ is considered and we show results for the isothermal pressure at the lowest temperature considered in this work. This choice of $T$ makes the PIMC simulation for densities in the deep quantum degenerate regime highly demanding in terms of the required number of imaginary time steps $M$ and calculations can be safely carried out only for relatively small system sizes. In Fig.~\ref{fig1}, the total number of particles is $N=216$ and at the highest density shown the required number of imaginary time steps is $M=160$. The comparison with the $T=0$ case is important to understand the effects of a finite temperature. In fact, at $T=0$ the pressure can be calculated using the energy functional in Ref.~\cite{PhysRevLett.115.155302} yielding the result
\begin{equation}
p=\frac{1}{4}gn^2\left(1+\frac{a_{12}}{a}\right)+\frac{4}{5\pi^2}\frac{m^{3/2}(gn)^{5/2}}{\hbar^3} \;,
\label{Petrov}
\end{equation}
in terms of the coupling constant $g=\frac{4\pi\hbar^2a}{m}$. The first term in the equation above is negative and corresponds to the mean-field instability while the second term is positive and provides the beyond mean-field stabilizing effect. As shown in Fig.~\ref{fig1}, the pressure in Eq.~(\ref{Petrov}) decreases with density, resulting in a negative inverse compressibility $\kappa_T^{-1}=n\frac{dp}{dn}$, then reaches a minimum, and finally it increases monotonically. The density at which the pressure crosses the $p=0$ value corresponds to the minimum of the energy functional and is the equilibrium density of the liquid phase reached in the center of large droplets. The corresponding gas parameter $na^3\sim10^{-4}$ agrees with the estimated value for the droplets observed in experiments~\cite{Science.359.301,PhysRevLett.120.235301,PhysRevResearch.1.033155}. This picture changes completely at $T>0$ where the liquid is no longer in equilibrium with the vacuum at zero pressure, but with a low density gas at some finite value of $p$. At finite $T$ an important density scale is the BEC density $n_{\mathrm{BEC}}=2\zeta(3/2)/\lambda_T^3$, where each component would undergo BEC in the absence of interactions (here $\lambda_T=\sqrt{2\pi\hbar^2/mk_BT}$ is the thermal wavelength and $\zeta(3/2)\simeq2.612$). For densities below $n_{\mathrm{BEC}}$ the mixture is in the gas phase and the pressure can be safely calculated using the Hartree-Fock scheme
\begin{equation}
p=\frac{1}{4}gn^2\left(2+\frac{a_{12}}{a}\right)+\frac{2k_BT}{\lambda_T^3}g_{5/2}(e^{\beta\tilde{\mu}}) \;,
\label{HF}
\end{equation}
where the factor 2 in the first term is due to exchange effects in the normal phase and the effective chemical potential $\tilde{\mu}$ is fixed by the normalization $n=2g_{3/2}(e^{\beta\tilde{\mu}})/\lambda_T^3$ in terms of the standard special Bose functions $g_\nu(z)$. It is important to stress that well-grounded approximate theories do not exist for $n>n_{\mathrm{BEC}}$: at the mean-field level one predicts a negative compressibility while beyond mean-field Bogoliubov approaches are plagued by a complex value of the speed of sound in the spectrum of elementary excitations~\footnote{At extremely low temperatures and close to the equilibrium density of the liquid at $T=0$, one can use the positive compressibility of the liquid state from Eq.~(\ref{Petrov}) to estimate the contribution to thermodynamics from phonon excitations as in Ref.~\cite{10.21468/SciPostPhys.9.2.020}}.

\begin{figure}
	\centering
	\includegraphics[width=0.85\columnwidth]{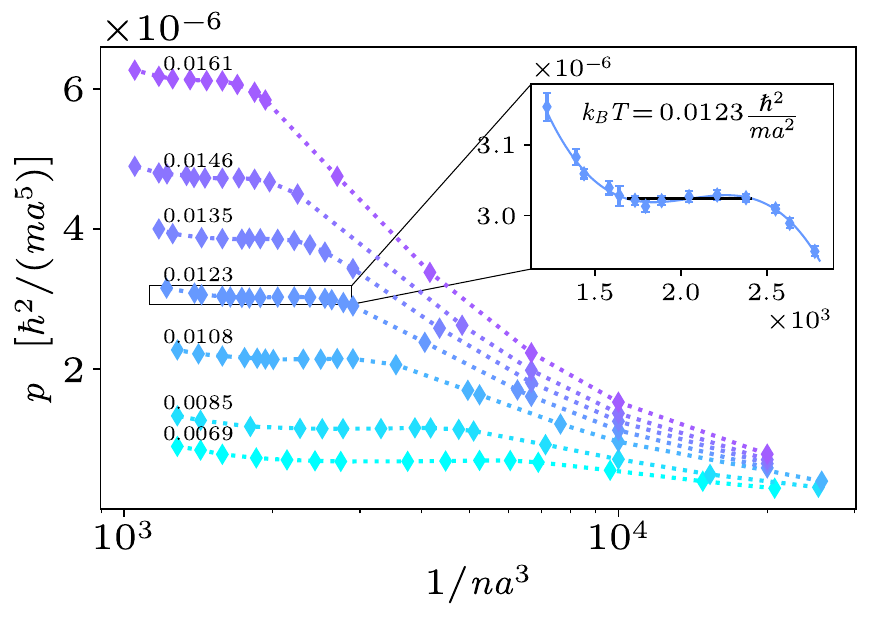}\vspace{-5pt}
	\caption{Isothermal curves of pressure as a function of density below and across the critical point. PIMC results refer to systems with $N=512$ particles and the dotted lines are a guide to the eye. Labels on each curve indicate the value of $k_B T$ in units of $\hbar^2/ma^2$. The inset shows the cubic fit to the points near the phase transition and the Maxwell construction used to determine the gap parameter $\Delta n$.}
	\label{fig2}
\end{figure}

Our PIMC results along the isothermal curve are in very good agreement with Eq.~(\ref{HF}) for $n<n_{\mathrm{BEC}}$, while for larger densities they feature a minimum, occurring at $p>0$, followed by a steady increase. A negative compressibility, corresponding to the region where $p$ decreases with $n$, is expected in a finite-$N$ system featuring coexisting phases as a result of the interface free energy~\cite{AllenTildesley}. The liquid-vapor coexistence region shall be pinpointed using the Maxwell construction with larger $N$, as discussed below. Still, we see that, already for $N=216$, the zero of pressure is lifted and there is no minimum in the free energy per particle $F/N$ at finite density.
One should point out that at any fixed temperature $T$, the entropy contribution yields a diverging $F/N=k_BT\log(n\lambda_T^3/2)$ for vanishing densities. Only at very low temperatures a local minimum in $F/N$ at finite density is present, indicating the existence of a metastable liquid state, whereby droplets are surrounded by vacuum. Estimates in Ref.~\cite{10.21468/SciPostPhys.9.2.020} show that this metastable state disappears for values of $T$ about $10$ times smaller than the one reported in Fig.~\ref{fig1}.
In our simulations, the liquid can exist in equilibrium with a finite-density gas, or in a homogeneous phase but at densities larger than the critical ones for phase coexistence, which are discussed later.
At the temperature considered in Fig.~\ref{fig1} we expect that for larger values of $N$ the pressure flattens near the finite-$N$ maximum. This value of pressure is expected to be close to $p_{\mathrm{BEC}}=\frac{gn_{\mathrm{BEC}}^2}{4}(2+a_{12}/a)+\frac{2k_BT}{\lambda_T^3}\zeta(5/2)$ (horizontal dashed line in Fig.~\ref{fig1}) corresponding to the pressure at the onset of BEC with $\zeta(5/2)\simeq1.341$.

A constant pressure in the isothermal curves signals the liquid-vapor coexistence region, which we obtain via the Maxwell construction for the larger values of $T$ reported in Fig.~\ref{fig2}. In this case we are able to perform simulations well converged in the number of imaginary time steps for significantly larger values of $N$.
The isothermal lines in Fig.~\ref{fig2} correspond to $N=512$ and still feature the typical S-like shape of finite-size systems in the coexistence region. Above the critical point the pressure is instead a monotonically increasing function of the density. The results for the pressure along a given isothermal line are fitted using a cubic curve in the volume per particle $1/n$ and the Maxwell construction is applied to determine the gap parameter $\Delta n=n_L-n_G$ between the densities respectively of the liquid and the gas. These two densities delimit the constant $p$ region. The Maxwell construction amounts to require the following condition on the fitting function $\int_{n_G^{-1}}^{n_L^{-1}}p\;d(1/n)=0$ (see inset of Fig.~\ref{fig2}). We also checked that, by increasing further the number of particles, the values of $n_L$ and $n_G$ extracted from the Maxwell construction do not change appreciably~\cite{Note4}.

\begin{table}[b]
    \begin{ruledtabular}
	\begin{tabular}{cccc}
		& $k_BT_c$ $[\hbar^2/ma^2]$ & $p_c$ $[\hbar^2/ma^5]$   & $n_ca^3$  \\[2pt]
		\hline
		$a_{12}=-1.2a$       & 0.0138(8)  & $4.1(6)\times 10^{-6}$ & $5.7(4)\times 10^{-4}$   \\
		$a_{12}=-1.1a$       & 0.0113(9)  & $2.5(5)\times 10^{-6}$  &  $4.2(4)\times 10^{-4}$    \\
	\end{tabular}
    \end{ruledtabular}
	\caption{Estimated values of the critical temperature, pressure and density for two values of $a_{12}$.}
    \label{tab1}
\end{table}

\begin{figure}
	\centering
	\includegraphics[width=0.85\columnwidth]{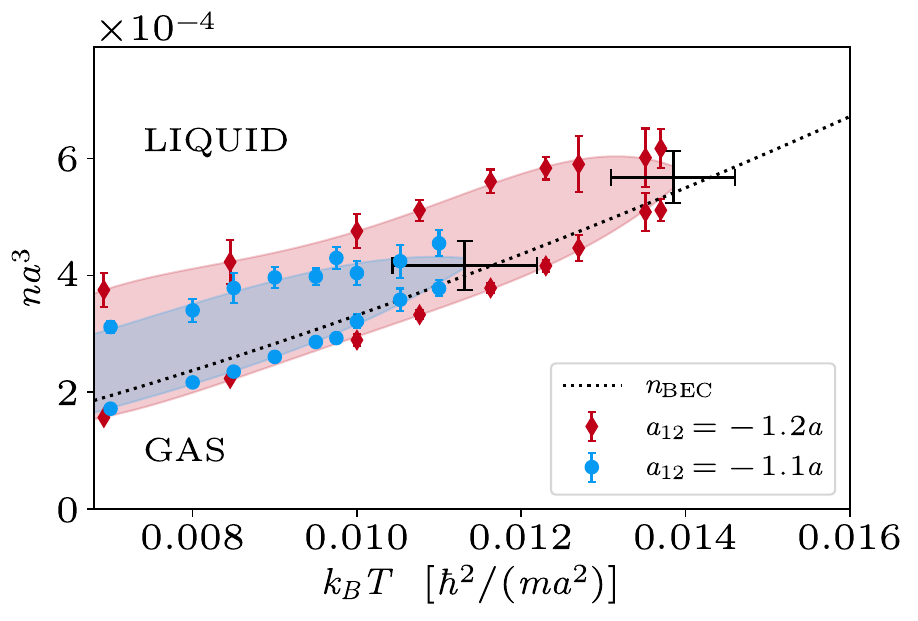}\vspace{-5pt}
	\caption{Phase diagram in the temperature vs density plane for two values of $a_{12}$. The shaded area corresponds to the liquid-gas coexistence region and the cross is our estimate (with uncertainties) of the critical point. The dotted line indicates the BEC transition density $n_{\mathrm{BEC}}=2\zeta(3/2)/\lambda_T^3$.}
	\label{fig3}
\end{figure}

\begin{figure}
	\centering
	\includegraphics[width=0.85\columnwidth]{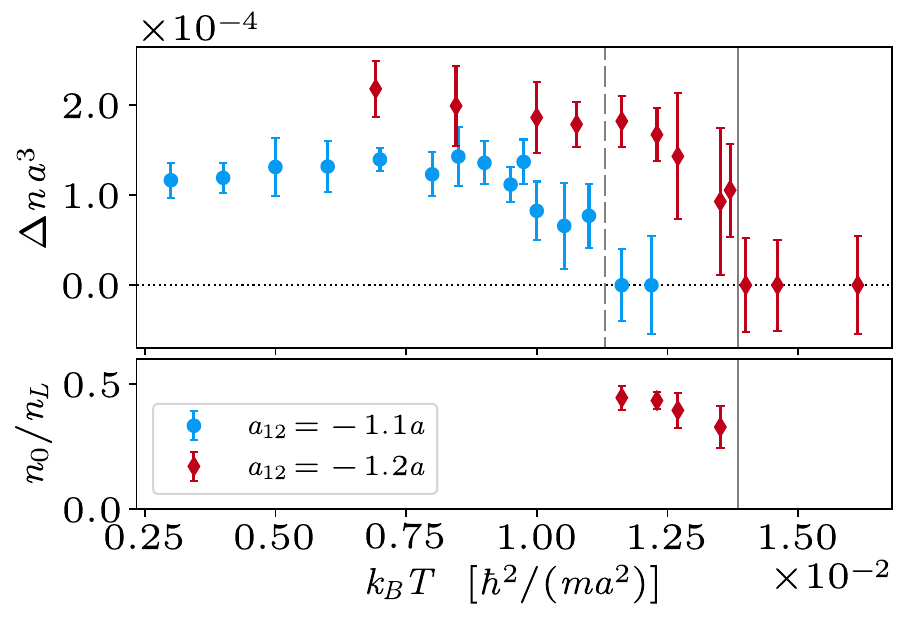}\vspace{-5pt}
	\caption{Decay of the order parameter $\Delta n$ (upper panel) and of the condensate fraction discontinuity (lower panel) with $T$ on approaching $T_c$. The vertical lines indicate our central value estimate of $T_c$.}
	\label{fig4}
\end{figure}

The results for the gap $\Delta n$ are reported in Fig.~\ref{fig3} for different temperatures and two values of the interspecies coupling $a_{12}$. The shaded region represents the coexistence between the two phases and the critical point is estimated from the isothermal line for which the Maxwell construction yields $\Delta n=0$. The results for the critical temperature $T_c$, pressure $p_c$, and density $n_c$ are reported in Table~\ref{tab1} for the two values of $a_{12}$~\cite{Note4}. Furthermore, we notice that the lower extreme $n_G$ of the coexistence region in Fig.~\ref{fig3} is slightly lower than the density $n_{\mathrm{BEC}}$, which signals the onset of BEC. As a result, the phases in the coexisting region involve a normal gas, on the verge of the BEC transition, and a superfluid liquid featuring a finite value of the condensate density $n_0$. Both the gap parameter $\Delta n$ and the discontinuous jump in the condensate density $n_0$ at the first-order vapor-liquid transition are reported in Fig.~\ref{fig4} as a function of temperature showing their closure at the critical temperature $T_c$. The condensate fraction $n_0/n_L$ is determined using the density $n_L$ of the homogeneous liquid phase at the upper extreme of the coexistence region~\footnote{The condensate density $n_0$ is computed via the one-body density matrix in PIMC simulations \cite{Note4}.}.

\begin{figure}
	\centering
	\includegraphics[width=0.85\columnwidth]{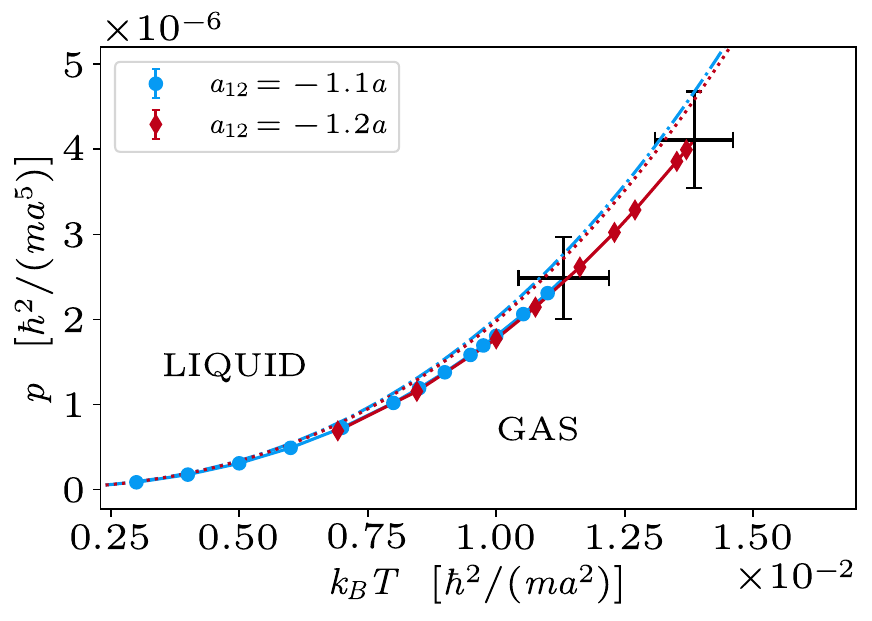}\vspace{-5pt}
	\caption{Phase diagram in the temperature vs pressure plane for two values of $a_{12}$. The solid line indicates the coexistence line between liquid and gas and the cross is our estimate (with uncertainties) of the critical point. The dotted lines refer to the pressure $p_{\mathrm{BEC}}$ at the onset of the BEC transition.}
	\label{fig5}
\end{figure}

The phase diagram in the temperature vs pressure plane, instead, is shown in Fig.~\ref{fig5}. The pressure along the coexistence line lies slightly below the pressure $p_{\mathrm{BEC}}$ of the gas at the onset of BEC. This finding is consistent with $n_G$, at the lower extreme of the coexistence region, being slightly smaller than $n_{\mathrm{BEC}}$ (see Fig.~\ref{fig3}). The coexistence lines for the two values of $a_{12}$ shown in Fig.~\ref{fig4} terminate at the corresponding critical points. For values of pressure and temperature larger respectively than $p_c$ and $T_c$ there is no difference between the liquid and the gas phase and one crosses over from one to the other in a continuous way.
The critical point is the ending point of the first-order transition line. For larger temperatures and pressures the BEC transition line is second order without density discontinuity.
We remark the peculiarity of the phase diagram's topology compared to the one of $^4$He and ultracold gases known so far. General aspects of the phase diagram in ultraquantum liquids and the possible emergence of a tricritical point have been discussed in Refs.~\cite{Son_2021, 10.1073/pnas.2017646117}.

A possible way to observe the liquid to gas first-order transition in attractive Bose mixtures is by investigating the density profile of the cloud confined in large harmonic traps $V_{\text{ext}}({\bf r})$ where the local density approximation can be safely applied in the form of the condition $\mu_{\text{local}}(T,n({\bf r}))+V_{\text{ext}}({\bf r})={\text{const}}$. By analyzing {\it in situ} the density as a function of the distance from the center of the cloud one should be able to observe the density jump $n_L-n_G$ at the distance corresponding to a local chemical potential $\mu_{\text{local}}=\mu_L(n_L)=\mu_G(n_G)$, where $\mu_L$ and $\mu_G$ are respectively the chemical potential of the liquid and the gas at the two densities $n_L$ and $n_G$. A similar procedure was used with polarized Fermi mixtures to observe the first-order transition between a paired superfluid and a polarized normal gas~\cite{PhysRevLett.101.070404,Shin2008}.
We also note that the possibility of characterizing the liquid-gas coexistence via the discontinuity of the density profile in trapped systems was put forward in Ref.~\cite{PhysRevA.107.L031303} considering ground-state Bose mixtures with Rabi coupling.
The discontinuous jump of the condensate density $n_0$ should also emerge from the analysis of the bimodal density distribution of the trapped cloud. Another possibility is to measure the canonical equation of state $p=p(T,n)$ where, according to the local density approximation, the system behaves locally as a homogeneous system at the density $n({\bf r})$. For a single component Bose gas the isothermal curves $p(T,n)$ were measured as a function of the density along the cloud profile crossing the BEC critical density~\cite{PhysRevLett.125.150404}. A similar experiment for the mixture should be well suited to capture the coexistence region where the pressure remains constant. Experiments on Bose mixtures carried out in box potentials with varying geometries~\cite{Navon2021} could also allow for access to the coexistence regime between liquid and vapor~\cite{PhysRevLett.130.193001}. It is also worth noticing that the use of heteronuclear mixtures could lead to much lower values of the critical density $n_c$ and consequently much longer lifetimes due to strong reduction of three-body losses~\cite{PhysRevResearch.1.033155}.

In conclusion, we used PIMC simulations to investigate the liquid to gas first-order transition in attractive binary Bose mixtures characterizing its critical point. The emerging picture shows intriguing analogies with the physics of classical fluids captured by Van der Waals theory of real gases. The mixture is, though, ultradilute and Bose condensed and the critical parameters are determined by quantum effects. Interesting new directions include the superfluid properties in the liquid phase, the physics of the liquid-gas interface and the study of asymmetric mixtures (different scattering lengths and different atomic masses) where a richer scenario of the liquid to gas transition is expected to occur.
The data presented in this Letter are freely available from Ref.~\cite{spada_2023_7845396}.\\

We acknowledge the Italian Ministry of University and Research under the PRIN2017 project CEnTraL 20172H2SC4.
G.S and S.G also acknowledge funding from the Provincia Autonoma di Trento.
S.P. acknowledges support from the PNRR MUR project PE0000023-NQSTI and from the CINECA awards IsCa6\_NEMCAQS and IsCb2\_NEMCASRA, for the availability of high performance computing resources and support.
G.S. acknowledges the CINECA award IsCa8\_QuaLiT under the ISCRA initiative, for the availability of high performance computing resources and support.
S.G. acknowledges support from ICSC – Centro Nazionale di Ricerca in HPC, Big Data and Quantum Computing, funded by the European Union under NextGenerationEU. Views and opinions expressed are, however, those of the author(s) only and do not necessarily reflect those of the European Union or The European Research Executive Agency. Neither the European Union nor the granting authority can be held responsible for them.

\bibliography{CriticalPoint}

\begin{thebibliography}{47}%
\makeatletter
\providecommand \@ifxundefined [1]{%
 \@ifx{#1\undefined}
}%
\providecommand \@ifnum [1]{%
 \ifnum #1\expandafter \@firstoftwo
 \else \expandafter \@secondoftwo
 \fi
}%
\providecommand \@ifx [1]{%
 \ifx #1\expandafter \@firstoftwo
 \else \expandafter \@secondoftwo
 \fi
}%
\providecommand \natexlab [1]{#1}%
\providecommand \enquote  [1]{``#1''}%
\providecommand \bibnamefont  [1]{#1}%
\providecommand \bibfnamefont [1]{#1}%
\providecommand \citenamefont [1]{#1}%
\providecommand \href@noop [0]{\@secondoftwo}%
\providecommand \href [0]{\begingroup \@sanitize@url \@href}%
\providecommand \@href[1]{\@@startlink{#1}\@@href}%
\providecommand \@@href[1]{\endgroup#1\@@endlink}%
\providecommand \@sanitize@url [0]{\catcode `\\12\catcode `\$12\catcode
  `\&12\catcode `\#12\catcode `\^12\catcode `\_12\catcode `\%12\relax}%
\providecommand \@@startlink[1]{}%
\providecommand \@@endlink[0]{}%
\providecommand \url  [0]{\begingroup\@sanitize@url \@url }%
\providecommand \@url [1]{\endgroup\@href {#1}{\urlprefix }}%
\providecommand \urlprefix  [0]{URL }%
\providecommand \Eprint [0]{\href }%
\providecommand \doibase [0]{https://doi.org/}%
\providecommand \selectlanguage [0]{\@gobble}%
\providecommand \bibinfo  [0]{\@secondoftwo}%
\providecommand \bibfield  [0]{\@secondoftwo}%
\providecommand \translation [1]{[#1]}%
\providecommand \BibitemOpen [0]{}%
\providecommand \bibitemStop [0]{}%
\providecommand \bibitemNoStop [0]{.\EOS\space}%
\providecommand \EOS [0]{\spacefactor3000\relax}%
\providecommand \BibitemShut  [1]{\csname bibitem#1\endcsname}%
\let\auto@bib@innerbib\@empty
\bibitem [{Note1()}]{Note1}%
  \BibitemOpen
  \bibinfo {note} {See {\protect \it e.g.} Ref.~\cite {book}
  Sec.~11.6}\BibitemShut {NoStop}%
\bibitem [{\citenamefont {Schmitt}\ \emph {et~al.}(2016)\citenamefont
  {Schmitt}, \citenamefont {Wenzel}, \citenamefont {B{\"o}ttcher},
  \citenamefont {Ferrier-Barbut},\ and\ \citenamefont {Pfau}}]{Schmitt2016}%
  \BibitemOpen
  \bibfield  {author} {\bibinfo {author} {\bibfnamefont {M.}~\bibnamefont
  {Schmitt}}, \bibinfo {author} {\bibfnamefont {M.}~\bibnamefont {Wenzel}},
  \bibinfo {author} {\bibfnamefont {F.}~\bibnamefont {B{\"o}ttcher}}, \bibinfo
  {author} {\bibfnamefont {I.}~\bibnamefont {Ferrier-Barbut}},\ and\ \bibinfo
  {author} {\bibfnamefont {T.}~\bibnamefont {Pfau}},\ }\bibfield  {title}
  {\bibinfo {title} {Self-bound droplets of a dilute magnetic quantum liquid},\
  }\href {https://doi.org/10.1038/nature20126} {\bibfield  {journal} {\bibinfo
  {journal} {Nature}\ }\textbf {\bibinfo {volume} {539}},\ \bibinfo {pages}
  {259} (\bibinfo {year} {2016})}\BibitemShut {NoStop}%
\bibitem [{\citenamefont {Chomaz}\ \emph {et~al.}(2016)\citenamefont {Chomaz},
  \citenamefont {Baier}, \citenamefont {Petter}, \citenamefont {Mark},
  \citenamefont {W\"achtler}, \citenamefont {Santos},\ and\ \citenamefont
  {Ferlaino}}]{PhysRevX.6.041039}%
  \BibitemOpen
  \bibfield  {author} {\bibinfo {author} {\bibfnamefont {L.}~\bibnamefont
  {Chomaz}}, \bibinfo {author} {\bibfnamefont {S.}~\bibnamefont {Baier}},
  \bibinfo {author} {\bibfnamefont {D.}~\bibnamefont {Petter}}, \bibinfo
  {author} {\bibfnamefont {M.~J.}\ \bibnamefont {Mark}}, \bibinfo {author}
  {\bibfnamefont {F.}~\bibnamefont {W\"achtler}}, \bibinfo {author}
  {\bibfnamefont {L.}~\bibnamefont {Santos}},\ and\ \bibinfo {author}
  {\bibfnamefont {F.}~\bibnamefont {Ferlaino}},\ }\bibfield  {title} {\bibinfo
  {title} {Quantum-fluctuation-driven crossover from a dilute {Bose-Einstein}
  condensate to a macrodroplet in a dipolar quantum fluid},\ }\href
  {https://doi.org/10.1103/PhysRevX.6.041039} {\bibfield  {journal} {\bibinfo
  {journal} {Phys. Rev. X}\ }\textbf {\bibinfo {volume} {6}},\ \bibinfo {pages}
  {041039} (\bibinfo {year} {2016})}\BibitemShut {NoStop}%
\bibitem [{\citenamefont {Cabrera}\ \emph {et~al.}(2018)\citenamefont
  {Cabrera}, \citenamefont {Tanzi}, \citenamefont {Sanz}, \citenamefont
  {Naylor}, \citenamefont {Thomas}, \citenamefont {Cheiney},\ and\
  \citenamefont {Tarruell}}]{Science.359.301}%
  \BibitemOpen
  \bibfield  {author} {\bibinfo {author} {\bibfnamefont {C.~R.}\ \bibnamefont
  {Cabrera}}, \bibinfo {author} {\bibfnamefont {L.}~\bibnamefont {Tanzi}},
  \bibinfo {author} {\bibfnamefont {J.}~\bibnamefont {Sanz}}, \bibinfo {author}
  {\bibfnamefont {B.}~\bibnamefont {Naylor}}, \bibinfo {author} {\bibfnamefont
  {P.}~\bibnamefont {Thomas}}, \bibinfo {author} {\bibfnamefont
  {P.}~\bibnamefont {Cheiney}},\ and\ \bibinfo {author} {\bibfnamefont
  {L.}~\bibnamefont {Tarruell}},\ }\bibfield  {title} {\bibinfo {title}
  {Quantum liquid droplets in a mixture of {Bose-Einstein} condensates},\
  }\href@noop {} {\bibfield  {journal} {\bibinfo  {journal} {Science}\ }\textbf
  {\bibinfo {volume} {359}},\ \bibinfo {pages} {301} (\bibinfo {year}
  {2018})}\BibitemShut {NoStop}%
\bibitem [{\citenamefont {Semeghini}\ \emph {et~al.}(2018)\citenamefont
  {Semeghini}, \citenamefont {Ferioli}, \citenamefont {Masi}, \citenamefont
  {Mazzinghi}, \citenamefont {Wolswijk}, \citenamefont {Minardi}, \citenamefont
  {Modugno}, \citenamefont {Modugno}, \citenamefont {Inguscio},\ and\
  \citenamefont {Fattori}}]{PhysRevLett.120.235301}%
  \BibitemOpen
  \bibfield  {author} {\bibinfo {author} {\bibfnamefont {G.}~\bibnamefont
  {Semeghini}}, \bibinfo {author} {\bibfnamefont {G.}~\bibnamefont {Ferioli}},
  \bibinfo {author} {\bibfnamefont {L.}~\bibnamefont {Masi}}, \bibinfo {author}
  {\bibfnamefont {C.}~\bibnamefont {Mazzinghi}}, \bibinfo {author}
  {\bibfnamefont {L.}~\bibnamefont {Wolswijk}}, \bibinfo {author}
  {\bibfnamefont {F.}~\bibnamefont {Minardi}}, \bibinfo {author} {\bibfnamefont
  {M.}~\bibnamefont {Modugno}}, \bibinfo {author} {\bibfnamefont
  {G.}~\bibnamefont {Modugno}}, \bibinfo {author} {\bibfnamefont
  {M.}~\bibnamefont {Inguscio}},\ and\ \bibinfo {author} {\bibfnamefont
  {M.}~\bibnamefont {Fattori}},\ }\bibfield  {title} {\bibinfo {title}
  {Self-bound quantum droplets of atomic mixtures in free space},\ }\href
  {https://doi.org/10.1103/PhysRevLett.120.235301} {\bibfield  {journal}
  {\bibinfo  {journal} {Phys. Rev. Lett.}\ }\textbf {\bibinfo {volume} {120}},\
  \bibinfo {pages} {235301} (\bibinfo {year} {2018})}\BibitemShut {NoStop}%
\bibitem [{\citenamefont {D'Errico}\ \emph {et~al.}(2019)\citenamefont
  {D'Errico}, \citenamefont {Burchianti}, \citenamefont {Prevedelli},
  \citenamefont {Salasnich}, \citenamefont {Ancilotto}, \citenamefont
  {Modugno}, \citenamefont {Minardi},\ and\ \citenamefont
  {Fort}}]{PhysRevResearch.1.033155}%
  \BibitemOpen
  \bibfield  {author} {\bibinfo {author} {\bibfnamefont {C.}~\bibnamefont
  {D'Errico}}, \bibinfo {author} {\bibfnamefont {A.}~\bibnamefont
  {Burchianti}}, \bibinfo {author} {\bibfnamefont {M.}~\bibnamefont
  {Prevedelli}}, \bibinfo {author} {\bibfnamefont {L.}~\bibnamefont
  {Salasnich}}, \bibinfo {author} {\bibfnamefont {F.}~\bibnamefont
  {Ancilotto}}, \bibinfo {author} {\bibfnamefont {M.}~\bibnamefont {Modugno}},
  \bibinfo {author} {\bibfnamefont {F.}~\bibnamefont {Minardi}},\ and\ \bibinfo
  {author} {\bibfnamefont {C.}~\bibnamefont {Fort}},\ }\bibfield  {title}
  {\bibinfo {title} {Observation of quantum droplets in a heteronuclear bosonic
  mixture},\ }\href {https://doi.org/10.1103/PhysRevResearch.1.033155}
  {\bibfield  {journal} {\bibinfo  {journal} {Phys. Rev. Res.}\ }\textbf
  {\bibinfo {volume} {1}},\ \bibinfo {pages} {033155} (\bibinfo {year}
  {2019})}\BibitemShut {NoStop}%
\bibitem [{\citenamefont {Petrov}(2015)}]{PhysRevLett.115.155302}%
  \BibitemOpen
  \bibfield  {author} {\bibinfo {author} {\bibfnamefont {D.~S.}\ \bibnamefont
  {Petrov}},\ }\bibfield  {title} {\bibinfo {title} {Quantum mechanical
  stabilization of a collapsing {Bose-Bose} mixture},\ }\href
  {https://doi.org/10.1103/PhysRevLett.115.155302} {\bibfield  {journal}
  {\bibinfo  {journal} {Phys. Rev. Lett.}\ }\textbf {\bibinfo {volume} {115}},\
  \bibinfo {pages} {155302} (\bibinfo {year} {2015})}\BibitemShut {NoStop}%
\bibitem [{\citenamefont {Macia}\ \emph {et~al.}(2016)\citenamefont {Macia},
  \citenamefont {S\'anchez-Baena}, \citenamefont {Boronat},\ and\ \citenamefont
  {Mazzanti}}]{PhysRevLett.117.205301}%
  \BibitemOpen
  \bibfield  {author} {\bibinfo {author} {\bibfnamefont {A.}~\bibnamefont
  {Macia}}, \bibinfo {author} {\bibfnamefont {J.}~\bibnamefont
  {S\'anchez-Baena}}, \bibinfo {author} {\bibfnamefont {J.}~\bibnamefont
  {Boronat}},\ and\ \bibinfo {author} {\bibfnamefont {F.}~\bibnamefont
  {Mazzanti}},\ }\bibfield  {title} {\bibinfo {title} {Droplets of trapped
  quantum dipolar bosons},\ }\href
  {https://doi.org/10.1103/PhysRevLett.117.205301} {\bibfield  {journal}
  {\bibinfo  {journal} {Phys. Rev. Lett.}\ }\textbf {\bibinfo {volume} {117}},\
  \bibinfo {pages} {205301} (\bibinfo {year} {2016})}\BibitemShut {NoStop}%
\bibitem [{\citenamefont {B\"ottcher}\ \emph {et~al.}(2019)\citenamefont
  {B\"ottcher}, \citenamefont {Wenzel}, \citenamefont {Schmidt}, \citenamefont
  {Guo}, \citenamefont {Langen}, \citenamefont {Ferrier-Barbut}, \citenamefont
  {Pfau}, \citenamefont {Bomb\'{\i}n}, \citenamefont {S\'anchez-Baena},
  \citenamefont {Boronat},\ and\ \citenamefont
  {Mazzanti}}]{PhysRevResearch.1.033088}%
  \BibitemOpen
  \bibfield  {author} {\bibinfo {author} {\bibfnamefont {F.}~\bibnamefont
  {B\"ottcher}}, \bibinfo {author} {\bibfnamefont {M.}~\bibnamefont {Wenzel}},
  \bibinfo {author} {\bibfnamefont {J.-N.}\ \bibnamefont {Schmidt}}, \bibinfo
  {author} {\bibfnamefont {M.}~\bibnamefont {Guo}}, \bibinfo {author}
  {\bibfnamefont {T.}~\bibnamefont {Langen}}, \bibinfo {author} {\bibfnamefont
  {I.}~\bibnamefont {Ferrier-Barbut}}, \bibinfo {author} {\bibfnamefont
  {T.}~\bibnamefont {Pfau}}, \bibinfo {author} {\bibfnamefont {R.}~\bibnamefont
  {Bomb\'{\i}n}}, \bibinfo {author} {\bibfnamefont {J.}~\bibnamefont
  {S\'anchez-Baena}}, \bibinfo {author} {\bibfnamefont {J.}~\bibnamefont
  {Boronat}},\ and\ \bibinfo {author} {\bibfnamefont {F.}~\bibnamefont
  {Mazzanti}},\ }\bibfield  {title} {\bibinfo {title} {Dilute dipolar quantum
  droplets beyond the extended {Gross-Pitaevskii} equation},\ }\href
  {https://doi.org/10.1103/PhysRevResearch.1.033088} {\bibfield  {journal}
  {\bibinfo  {journal} {Phys. Rev. Res.}\ }\textbf {\bibinfo {volume} {1}},\
  \bibinfo {pages} {033088} (\bibinfo {year} {2019})}\BibitemShut {NoStop}%
\bibitem [{\citenamefont {Cikojevi\ifmmode~\acute{c}\else \'{c}\fi{}}\ \emph
  {et~al.}(2018)\citenamefont {Cikojevi\ifmmode~\acute{c}\else \'{c}\fi{}},
  \citenamefont {D\ifmmode~\check{z}\else \v{z}\fi{}elalija}, \citenamefont
  {Stipanovi\ifmmode~\acute{c}\else \'{c}\fi{}}, \citenamefont {Vranje\ifmmode
  \check{s}\else \v{s}\fi{} Marki\ifmmode~\acute{c}\else \'{c}\fi{}},\ and\
  \citenamefont {Boronat}}]{PhysRevB.97.140502}%
  \BibitemOpen
  \bibfield  {author} {\bibinfo {author} {\bibfnamefont {V.}~\bibnamefont
  {Cikojevi\ifmmode~\acute{c}\else \'{c}\fi{}}}, \bibinfo {author}
  {\bibfnamefont {K.}~\bibnamefont {D\ifmmode~\check{z}\else
  \v{z}\fi{}elalija}}, \bibinfo {author} {\bibfnamefont {P.}~\bibnamefont
  {Stipanovi\ifmmode~\acute{c}\else \'{c}\fi{}}}, \bibinfo {author}
  {\bibfnamefont {L.}~\bibnamefont {Vranje\ifmmode \check{s}\else \v{s}\fi{}
  Marki\ifmmode~\acute{c}\else \'{c}\fi{}}},\ and\ \bibinfo {author}
  {\bibfnamefont {J.}~\bibnamefont {Boronat}},\ }\bibfield  {title} {\bibinfo
  {title} {Ultradilute quantum liquid drops},\ }\href
  {https://doi.org/10.1103/PhysRevB.97.140502} {\bibfield  {journal} {\bibinfo
  {journal} {Phys. Rev. B}\ }\textbf {\bibinfo {volume} {97}},\ \bibinfo
  {pages} {140502} (\bibinfo {year} {2018})}\BibitemShut {NoStop}%
\bibitem [{\citenamefont {Cikojevi\ifmmode~\acute{c}\else \'{c}\fi{}}\ \emph
  {et~al.}(2019)\citenamefont {Cikojevi\ifmmode~\acute{c}\else \'{c}\fi{}},
  \citenamefont {Vranje\ifmmode \check{s}\else \v{s}\fi{}
  Marki\ifmmode~\acute{c}\else \'{c}\fi{}}, \citenamefont {Astrakharchik},\
  and\ \citenamefont {Boronat}}]{PhysRevA.99.023618}%
  \BibitemOpen
  \bibfield  {author} {\bibinfo {author} {\bibfnamefont {V.}~\bibnamefont
  {Cikojevi\ifmmode~\acute{c}\else \'{c}\fi{}}}, \bibinfo {author}
  {\bibfnamefont {L.}~\bibnamefont {Vranje\ifmmode \check{s}\else \v{s}\fi{}
  Marki\ifmmode~\acute{c}\else \'{c}\fi{}}}, \bibinfo {author} {\bibfnamefont
  {G.~E.}\ \bibnamefont {Astrakharchik}},\ and\ \bibinfo {author}
  {\bibfnamefont {J.}~\bibnamefont {Boronat}},\ }\bibfield  {title} {\bibinfo
  {title} {Universality in ultradilute liquid {Bose-Bose} mixtures},\ }\href
  {https://doi.org/10.1103/PhysRevA.99.023618} {\bibfield  {journal} {\bibinfo
  {journal} {Phys. Rev. A}\ }\textbf {\bibinfo {volume} {99}},\ \bibinfo
  {pages} {023618} (\bibinfo {year} {2019})}\BibitemShut {NoStop}%
\bibitem [{\citenamefont {Cikojevi\ifmmode~\acute{c}\else \'{c}\fi{}}\ \emph
  {et~al.}(2021)\citenamefont {Cikojevi\ifmmode~\acute{c}\else \'{c}\fi{}},
  \citenamefont {Poli}, \citenamefont {Ancilotto}, \citenamefont
  {Vranje\ifmmode \check{s}\else \v{s}\fi{} Marki\ifmmode~\acute{c}\else
  \'{c}\fi{}},\ and\ \citenamefont {Boronat}}]{PhysRevA.104.033319}%
  \BibitemOpen
  \bibfield  {author} {\bibinfo {author} {\bibfnamefont {V.}~\bibnamefont
  {Cikojevi\ifmmode~\acute{c}\else \'{c}\fi{}}}, \bibinfo {author}
  {\bibfnamefont {E.}~\bibnamefont {Poli}}, \bibinfo {author} {\bibfnamefont
  {F.}~\bibnamefont {Ancilotto}}, \bibinfo {author} {\bibfnamefont
  {L.}~\bibnamefont {Vranje\ifmmode \check{s}\else \v{s}\fi{}
  Marki\ifmmode~\acute{c}\else \'{c}\fi{}}},\ and\ \bibinfo {author}
  {\bibfnamefont {J.}~\bibnamefont {Boronat}},\ }\bibfield  {title} {\bibinfo
  {title} {Dilute quantum liquid in a {K-Rb} {Bose} mixture},\ }\href
  {https://doi.org/10.1103/PhysRevA.104.033319} {\bibfield  {journal} {\bibinfo
   {journal} {Phys. Rev. A}\ }\textbf {\bibinfo {volume} {104}},\ \bibinfo
  {pages} {033319} (\bibinfo {year} {2021})}\BibitemShut {NoStop}%
\bibitem [{\citenamefont {Petrov}\ and\ \citenamefont
  {Astrakharchik}(2016)}]{PhysRevLett.117.100401}%
  \BibitemOpen
  \bibfield  {author} {\bibinfo {author} {\bibfnamefont {D.~S.}\ \bibnamefont
  {Petrov}}\ and\ \bibinfo {author} {\bibfnamefont {G.~E.}\ \bibnamefont
  {Astrakharchik}},\ }\bibfield  {title} {\bibinfo {title} {Ultradilute
  low-dimensional liquids},\ }\href
  {https://doi.org/10.1103/PhysRevLett.117.100401} {\bibfield  {journal}
  {\bibinfo  {journal} {Phys. Rev. Lett.}\ }\textbf {\bibinfo {volume} {117}},\
  \bibinfo {pages} {100401} (\bibinfo {year} {2016})}\BibitemShut {NoStop}%
\bibitem [{\citenamefont {Parisi}\ \emph {et~al.}(2019)\citenamefont {Parisi},
  \citenamefont {Astrakharchik},\ and\ \citenamefont
  {Giorgini}}]{PhysRevLett.122.105302}%
  \BibitemOpen
  \bibfield  {author} {\bibinfo {author} {\bibfnamefont {L.}~\bibnamefont
  {Parisi}}, \bibinfo {author} {\bibfnamefont {G.~E.}\ \bibnamefont
  {Astrakharchik}},\ and\ \bibinfo {author} {\bibfnamefont {S.}~\bibnamefont
  {Giorgini}},\ }\bibfield  {title} {\bibinfo {title} {Liquid state of
  one-dimensional {Bose} mixtures: {A} quantum {Monte Carlo} study},\ }\href
  {https://doi.org/10.1103/PhysRevLett.122.105302} {\bibfield  {journal}
  {\bibinfo  {journal} {Phys. Rev. Lett.}\ }\textbf {\bibinfo {volume} {122}},\
  \bibinfo {pages} {105302} (\bibinfo {year} {2019})}\BibitemShut {NoStop}%
\bibitem [{\citenamefont {Parisi}\ and\ \citenamefont
  {Giorgini}(2020)}]{PhysRevA.102.023318}%
  \BibitemOpen
  \bibfield  {author} {\bibinfo {author} {\bibfnamefont {L.}~\bibnamefont
  {Parisi}}\ and\ \bibinfo {author} {\bibfnamefont {S.}~\bibnamefont
  {Giorgini}},\ }\bibfield  {title} {\bibinfo {title} {Quantum droplets in
  one-dimensional {Bose} mixtures: A quantum {Monte Carlo} study},\ }\href
  {https://doi.org/10.1103/PhysRevA.102.023318} {\bibfield  {journal} {\bibinfo
   {journal} {Phys. Rev. A}\ }\textbf {\bibinfo {volume} {102}},\ \bibinfo
  {pages} {023318} (\bibinfo {year} {2020})}\BibitemShut {NoStop}%
\bibitem [{\citenamefont {Saito}(2016)}]{doi:10.7566/JPSJ.85.053001}%
  \BibitemOpen
  \bibfield  {author} {\bibinfo {author} {\bibfnamefont {H.}~\bibnamefont
  {Saito}},\ }\bibfield  {title} {\bibinfo {title} {Path-integral {Monte Carlo}
  study on a droplet of a dipolar {Bose–Einstein} condensate stabilized by
  quantum fluctuation},\ }\href {https://doi.org/10.7566/JPSJ.85.053001}
  {\bibfield  {journal} {\bibinfo  {journal} {J. Phys. Soc. Jpn.}\ }\textbf
  {\bibinfo {volume} {85}},\ \bibinfo {pages} {053001} (\bibinfo {year}
  {2016})}\BibitemShut {NoStop}%
\bibitem [{\citenamefont {Ota}\ and\ \citenamefont
  {Astrakharchik}(2020)}]{10.21468/SciPostPhys.9.2.020}%
  \BibitemOpen
  \bibfield  {author} {\bibinfo {author} {\bibfnamefont {M.}~\bibnamefont
  {Ota}}\ and\ \bibinfo {author} {\bibfnamefont {G.~E.}\ \bibnamefont
  {Astrakharchik}},\ }\bibfield  {title} {\bibinfo {title} {{Beyond
  Lee-Huang-Yang description of self-bound Bose mixtures}},\ }\href
  {https://doi.org/10.21468/SciPostPhys.9.2.020} {\bibfield  {journal}
  {\bibinfo  {journal} {SciPost Phys.}\ }\textbf {\bibinfo {volume} {9}},\
  \bibinfo {pages} {020} (\bibinfo {year} {2020})}\BibitemShut {NoStop}%
\bibitem [{\citenamefont {Wang}\ \emph {et~al.}(2020)\citenamefont {Wang},
  \citenamefont {Hu},\ and\ \citenamefont {Liu}}]{Wang_2020}%
  \BibitemOpen
  \bibfield  {author} {\bibinfo {author} {\bibfnamefont {J.}~\bibnamefont
  {Wang}}, \bibinfo {author} {\bibfnamefont {H.}~\bibnamefont {Hu}},\ and\
  \bibinfo {author} {\bibfnamefont {X.-J.}\ \bibnamefont {Liu}},\ }\bibfield
  {title} {\bibinfo {title} {Thermal destabilization of self-bound ultradilute
  quantum droplets},\ }\href {https://doi.org/10.1088/1367-2630/abbe55}
  {\bibfield  {journal} {\bibinfo  {journal} {New J. Phys.}\ }\textbf {\bibinfo
  {volume} {22}},\ \bibinfo {pages} {103044} (\bibinfo {year}
  {2020})}\BibitemShut {NoStop}%
\bibitem [{\citenamefont {De~Rosi}\ \emph {et~al.}(2021)\citenamefont
  {De~Rosi}, \citenamefont {Astrakharchik},\ and\ \citenamefont
  {Massignan}}]{PhysRevA.103.043316}%
  \BibitemOpen
  \bibfield  {author} {\bibinfo {author} {\bibfnamefont {G.}~\bibnamefont
  {De~Rosi}}, \bibinfo {author} {\bibfnamefont {G.~E.}\ \bibnamefont
  {Astrakharchik}},\ and\ \bibinfo {author} {\bibfnamefont {P.}~\bibnamefont
  {Massignan}},\ }\bibfield  {title} {\bibinfo {title} {Thermal instability,
  evaporation, and thermodynamics of one-dimensional liquids in weakly
  interacting {Bose-Bose} mixtures},\ }\href
  {https://doi.org/10.1103/PhysRevA.103.043316} {\bibfield  {journal} {\bibinfo
   {journal} {Phys. Rev. A}\ }\textbf {\bibinfo {volume} {103}},\ \bibinfo
  {pages} {043316} (\bibinfo {year} {2021})}\BibitemShut {NoStop}%
\bibitem [{Note2()}]{Note2}%
  \BibitemOpen
  \bibinfo {note} {Notice that one-dimensional mixtures do not suffer from this
  shortcoming~\cite {PhysRevA.103.043316}}\BibitemShut {NoStop}%
\bibitem [{Note3()}]{Note3}%
  \BibitemOpen
  \bibinfo {note} {We note that liquid-gas coexistence states have been
  predicted at $T=0$ in mixtures with spin imbalance or coherent coupling in
  Refs~\cite {PhysRevA.107.L031303} and~\cite {he2022quantum}}\BibitemShut
  {NoStop}%
\bibitem [{\citenamefont {Gu}\ and\ \citenamefont
  {Cui}(2023)}]{PhysRevA.107.L031303}%
  \BibitemOpen
  \bibfield  {author} {\bibinfo {author} {\bibfnamefont {Q.}~\bibnamefont
  {Gu}}\ and\ \bibinfo {author} {\bibfnamefont {X.}~\bibnamefont {Cui}},\
  }\bibfield  {title} {\bibinfo {title} {Liquid-gas transition and coexistence
  in ground-state bosons with spin twist},\ }\href
  {https://doi.org/10.1103/PhysRevA.107.L031303} {\bibfield  {journal}
  {\bibinfo  {journal} {Phys. Rev. A}\ }\textbf {\bibinfo {volume} {107}},\
  \bibinfo {pages} {L031303} (\bibinfo {year} {2023})}\BibitemShut {NoStop}%
\bibitem [{\citenamefont {He}\ \emph {et~al.}(2023{\natexlab{a}})\citenamefont
  {He}, \citenamefont {Li}, \citenamefont {Yi},\ and\ \citenamefont
  {Yu}}]{he2022quantum}%
  \BibitemOpen
  \bibfield  {author} {\bibinfo {author} {\bibfnamefont {L.}~\bibnamefont
  {He}}, \bibinfo {author} {\bibfnamefont {H.}~\bibnamefont {Li}}, \bibinfo
  {author} {\bibfnamefont {W.}~\bibnamefont {Yi}},\ and\ \bibinfo {author}
  {\bibfnamefont {Z.-Q.}\ \bibnamefont {Yu}},\ }\bibfield  {title} {\bibinfo
  {title} {Quantum criticality of liquid-gas transition in a binary {Bose}
  mixture},\ }\href {https://doi.org/10.1103/PhysRevLett.130.193001} {\bibfield
   {journal} {\bibinfo  {journal} {Phys. Rev. Lett.}\ }\textbf {\bibinfo
  {volume} {130}},\ \bibinfo {pages} {193001} (\bibinfo {year}
  {2023}{\natexlab{a}})}\BibitemShut {NoStop}%
\bibitem [{\citenamefont {Spada}\ \emph
  {et~al.}(2022{\natexlab{a}})\citenamefont {Spada}, \citenamefont {Giorgini},\
  and\ \citenamefont {Pilati}}]{condmat7020030}%
  \BibitemOpen
  \bibfield  {author} {\bibinfo {author} {\bibfnamefont {G.}~\bibnamefont
  {Spada}}, \bibinfo {author} {\bibfnamefont {S.}~\bibnamefont {Giorgini}},\
  and\ \bibinfo {author} {\bibfnamefont {S.}~\bibnamefont {Pilati}},\
  }\bibfield  {title} {\bibinfo {title} {Path-integral {Monte Carlo} worm
  algorithm for {Bose} systems with periodic boundary conditions},\ }\href
  {https://doi.org/10.3390/condmat7020030} {\bibfield  {journal} {\bibinfo
  {journal} {Condens. Matter}\ }\textbf {\bibinfo {volume} {7}},\ \bibinfo
  {pages} {30} (\bibinfo {year} {2022}{\natexlab{a}})}\BibitemShut {NoStop}%
\bibitem [{Note4()}]{Note4}%
  \BibitemOpen
  \bibinfo {note} {See Supplemental Material, which includes Refs.~\cite
  {RevModPhys.67.279,PhysRevA.105.013325,PhysRevLett.96.070601,doi:10.1063/1.463076,Alder1962dis,reif2010history,PhysRevLett.29.28,10.1063/1.4720089,Rovere_1990}}\BibitemShut
  {NoStop}%
\bibitem [{\citenamefont {Ceperley}(1995)}]{RevModPhys.67.279}%
  \BibitemOpen
  \bibfield  {author} {\bibinfo {author} {\bibfnamefont {D.~M.}\ \bibnamefont
  {Ceperley}},\ }\bibfield  {title} {\bibinfo {title} {{Path integrals in the
  theory of condensed helium}},\ }\href
  {https://doi.org/10.1103/RevModPhys.67.279} {\bibfield  {journal} {\bibinfo
  {journal} {Rev. Mod. Phys.}\ }\textbf {\bibinfo {volume} {67}},\ \bibinfo
  {pages} {279} (\bibinfo {year} {1995})}\BibitemShut {NoStop}%
\bibitem [{\citenamefont {Spada}\ \emph
  {et~al.}(2022{\natexlab{b}})\citenamefont {Spada}, \citenamefont {Pilati},\
  and\ \citenamefont {Giorgini}}]{PhysRevA.105.013325}%
  \BibitemOpen
  \bibfield  {author} {\bibinfo {author} {\bibfnamefont {G.}~\bibnamefont
  {Spada}}, \bibinfo {author} {\bibfnamefont {S.}~\bibnamefont {Pilati}},\ and\
  \bibinfo {author} {\bibfnamefont {S.}~\bibnamefont {Giorgini}},\ }\bibfield
  {title} {\bibinfo {title} {Thermodynamics of a dilute {Bose} gas: A
  path-integral {Monte Carlo} study},\ }\href
  {https://doi.org/10.1103/PhysRevA.105.013325} {\bibfield  {journal} {\bibinfo
   {journal} {Phys. Rev. A}\ }\textbf {\bibinfo {volume} {105}},\ \bibinfo
  {pages} {013325} (\bibinfo {year} {2022}{\natexlab{b}})}\BibitemShut
  {NoStop}%
\bibitem [{\citenamefont {Boninsegni}\ \emph {et~al.}(2006)\citenamefont
  {Boninsegni}, \citenamefont {Prokof'ev},\ and\ \citenamefont
  {Svistunov}}]{PhysRevLett.96.070601}%
  \BibitemOpen
  \bibfield  {author} {\bibinfo {author} {\bibfnamefont {M.}~\bibnamefont
  {Boninsegni}}, \bibinfo {author} {\bibfnamefont {N.}~\bibnamefont
  {Prokof'ev}},\ and\ \bibinfo {author} {\bibfnamefont {B.}~\bibnamefont
  {Svistunov}},\ }\bibfield  {title} {\bibinfo {title} {{Worm algorithm for
  continuous-space path integral Monte Carlo simulations}},\ }\href
  {https://doi.org/10.1103/PhysRevLett.96.070601} {\bibfield  {journal}
  {\bibinfo  {journal} {Phys. Rev. Lett.}\ }\textbf {\bibinfo {volume} {96}},\
  \bibinfo {pages} {070601} (\bibinfo {year} {2006})}\BibitemShut {NoStop}%
\bibitem [{\citenamefont {Cao}\ and\ \citenamefont
  {Berne}(1992)}]{doi:10.1063/1.463076}%
  \BibitemOpen
  \bibfield  {author} {\bibinfo {author} {\bibfnamefont {J.}~\bibnamefont
  {Cao}}\ and\ \bibinfo {author} {\bibfnamefont {B.~J.}\ \bibnamefont
  {Berne}},\ }\bibfield  {title} {\bibinfo {title} {{A new quantum propagator
  for hard sphere and cavity systems}},\ }\href
  {https://doi.org/10.1063/1.463076} {\bibfield  {journal} {\bibinfo  {journal}
  {J. Chem. Phys.}\ }\textbf {\bibinfo {volume} {97}},\ \bibinfo {pages} {2382}
  (\bibinfo {year} {1992})}\BibitemShut {NoStop}%
\bibitem [{\citenamefont {Alder}\ and\ \citenamefont
  {Wainwright}(1962)}]{Alder1962dis}%
  \BibitemOpen
  \bibfield  {author} {\bibinfo {author} {\bibfnamefont {B.~J.}\ \bibnamefont
  {Alder}}\ and\ \bibinfo {author} {\bibfnamefont {T.~E.}\ \bibnamefont
  {Wainwright}},\ }\bibfield  {title} {\bibinfo {title} {Phase transition in
  elastic disks},\ }\href {https://doi.org/10.1103/PhysRev.127.359} {\bibfield
  {journal} {\bibinfo  {journal} {Phys. Rev.}\ }\textbf {\bibinfo {volume}
  {127}},\ \bibinfo {pages} {359} (\bibinfo {year} {1962})}\BibitemShut
  {NoStop}%
\bibitem [{\citenamefont {Reif-Acherman}(2010)}]{reif2010history}%
  \BibitemOpen
  \bibfield  {author} {\bibinfo {author} {\bibfnamefont {S.}~\bibnamefont
  {Reif-Acherman}},\ }\bibfield  {title} {\bibinfo {title} {The history of the
  rectilinear diameter law},\ }\href@noop {} {\bibfield  {journal} {\bibinfo
  {journal} {Quim. Nova}\ }\textbf {\bibinfo {volume} {33}},\ \bibinfo {pages}
  {2003} (\bibinfo {year} {2010})}\BibitemShut {NoStop}%
\bibitem [{\citenamefont {Cornfeld}\ and\ \citenamefont
  {Carr}(1972)}]{PhysRevLett.29.28}%
  \BibitemOpen
  \bibfield  {author} {\bibinfo {author} {\bibfnamefont {A.~B.}\ \bibnamefont
  {Cornfeld}}\ and\ \bibinfo {author} {\bibfnamefont {H.~Y.}\ \bibnamefont
  {Carr}},\ }\bibfield  {title} {\bibinfo {title} {Experimental evidence
  concerning the law of rectilinear diameter},\ }\href
  {https://doi.org/10.1103/PhysRevLett.29.28} {\bibfield  {journal} {\bibinfo
  {journal} {Phys. Rev. Lett.}\ }\textbf {\bibinfo {volume} {29}},\ \bibinfo
  {pages} {28} (\bibinfo {year} {1972})}\BibitemShut {NoStop}%
\bibitem [{\citenamefont {Watanabe}\ \emph {et~al.}(2012)\citenamefont
  {Watanabe}, \citenamefont {Ito},\ and\ \citenamefont
  {Hu}}]{10.1063/1.4720089}%
  \BibitemOpen
  \bibfield  {author} {\bibinfo {author} {\bibfnamefont {H.}~\bibnamefont
  {Watanabe}}, \bibinfo {author} {\bibfnamefont {N.}~\bibnamefont {Ito}},\ and\
  \bibinfo {author} {\bibfnamefont {C.-K.}\ \bibnamefont {Hu}},\ }\bibfield
  {title} {\bibinfo {title} {{Phase diagram and universality of the
  {Lennard-Jones} gas-liquid system}},\ }\bibfield  {journal} {\bibinfo
  {journal} {J. Chem. Phys.}\ }\textbf {\bibinfo {volume} {136}},\ \href
  {https://doi.org/10.1063/1.4720089} {10.1063/1.4720089} (\bibinfo {year}
  {2012}),\ \bibinfo {note} {204102}\BibitemShut {NoStop}%
\bibitem [{\citenamefont {Rovere}\ \emph {et~al.}(1990)\citenamefont {Rovere},
  \citenamefont {Heermann},\ and\ \citenamefont {Binder}}]{Rovere_1990}%
  \BibitemOpen
  \bibfield  {author} {\bibinfo {author} {\bibfnamefont {M.}~\bibnamefont
  {Rovere}}, \bibinfo {author} {\bibfnamefont {D.~W.}\ \bibnamefont
  {Heermann}},\ and\ \bibinfo {author} {\bibfnamefont {K.}~\bibnamefont
  {Binder}},\ }\bibfield  {title} {\bibinfo {title} {The gas-liquid transition
  of the two-dimensional {Lennard-Jones} fluid},\ }\href
  {https://doi.org/10.1088/0953-8984/2/33/013} {\bibfield  {journal} {\bibinfo
  {journal} {Journal of Physics: Condensed Matter}\ }\textbf {\bibinfo {volume}
  {2}},\ \bibinfo {pages} {7009} (\bibinfo {year} {1990})}\BibitemShut
  {NoStop}%
\bibitem [{\citenamefont {Pilati}\ \emph {et~al.}(2006)\citenamefont {Pilati},
  \citenamefont {Sakkos}, \citenamefont {Boronat}, \citenamefont {Casulleras},\
  and\ \citenamefont {Giorgini}}]{PhysRevA.74.043621}%
  \BibitemOpen
  \bibfield  {author} {\bibinfo {author} {\bibfnamefont {S.}~\bibnamefont
  {Pilati}}, \bibinfo {author} {\bibfnamefont {K.}~\bibnamefont {Sakkos}},
  \bibinfo {author} {\bibfnamefont {J.}~\bibnamefont {Boronat}}, \bibinfo
  {author} {\bibfnamefont {J.}~\bibnamefont {Casulleras}},\ and\ \bibinfo
  {author} {\bibfnamefont {S.}~\bibnamefont {Giorgini}},\ }\bibfield  {title}
  {\bibinfo {title} {{Equation of state of an interacting Bose gas at finite
  temperature: A path-integral Monte Carlo study}},\ }\href
  {https://doi.org/10.1103/PhysRevA.74.043621} {\bibfield  {journal} {\bibinfo
  {journal} {Phys. Rev. A}\ }\textbf {\bibinfo {volume} {74}},\ \bibinfo
  {pages} {043621} (\bibinfo {year} {2006})}\BibitemShut {NoStop}%
\bibitem [{Note5()}]{Note5}%
  \BibitemOpen
  \bibinfo {note} {At extremely low temperatures and close to the equilibrium
  density of the liquid at $T=0$, one can use the positive compressibility of
  the liquid state from Eq.~(\ref {Petrov}) to estimate the contribution to
  thermodynamics from phonon excitations as in Ref.~\cite
  {10.21468/SciPostPhys.9.2.020}}\BibitemShut {NoStop}%
\bibitem [{\citenamefont {Allen}\ and\ \citenamefont
  {Tildesley}(1987)}]{AllenTildesley}%
  \BibitemOpen
  \bibfield  {author} {\bibinfo {author} {\bibfnamefont {M.~P.}\ \bibnamefont
  {Allen}}\ and\ \bibinfo {author} {\bibfnamefont {D.~J.}\ \bibnamefont
  {Tildesley}},\ }\href@noop {} {\emph {\bibinfo {title} {Computer Simulations
  of Liquids}}}\ (\bibinfo  {publisher} {Oxford University Press, New York},\
  \bibinfo {year} {1987})\BibitemShut {NoStop}%
\bibitem [{Note6()}]{Note6}%
  \BibitemOpen
  \bibinfo {note} {The condensate density $n_0$ is computed via the one-body
  density matrix in PIMC simulations \cite {Note4}.}\BibitemShut {Stop}%
\bibitem [{\citenamefont {Son}\ \emph {et~al.}(2021)\citenamefont {Son},
  \citenamefont {Stephanov},\ and\ \citenamefont {Yee}}]{Son_2021}%
  \BibitemOpen
  \bibfield  {author} {\bibinfo {author} {\bibfnamefont {D.~T.}\ \bibnamefont
  {Son}}, \bibinfo {author} {\bibfnamefont {M.}~\bibnamefont {Stephanov}},\
  and\ \bibinfo {author} {\bibfnamefont {H.-U.}\ \bibnamefont {Yee}},\
  }\bibfield  {title} {\bibinfo {title} {The phase diagram of ultra quantum
  liquids},\ }\href {https://doi.org/10.1088/1742-5468/abd024} {\bibfield
  {journal} {\bibinfo  {journal} {J. Stat. Mech.}\ }\textbf {\bibinfo {volume}
  {2021}},\ \bibinfo {pages} {013105} (\bibinfo {year} {2021})}\BibitemShut
  {NoStop}%
\bibitem [{\citenamefont {Kora}\ \emph {et~al.}(2020)\citenamefont {Kora},
  \citenamefont {Boninsegni}, \citenamefont {Son},\ and\ \citenamefont
  {Zhang}}]{10.1073/pnas.2017646117}%
  \BibitemOpen
  \bibfield  {author} {\bibinfo {author} {\bibfnamefont {Y.}~\bibnamefont
  {Kora}}, \bibinfo {author} {\bibfnamefont {M.}~\bibnamefont {Boninsegni}},
  \bibinfo {author} {\bibfnamefont {D.~T.}\ \bibnamefont {Son}},\ and\ \bibinfo
  {author} {\bibfnamefont {S.}~\bibnamefont {Zhang}},\ }\bibfield  {title}
  {\bibinfo {title} {Tuning the quantumness of simple bose systems: A universal
  phase diagram},\ }\href {https://doi.org/10.1073/pnas.2017646117} {\bibfield
  {journal} {\bibinfo  {journal} {Proc. Natl. Acad. Sci. U.S.A.}\ }\textbf
  {\bibinfo {volume} {117}},\ \bibinfo {pages} {27231} (\bibinfo {year}
  {2020})}\BibitemShut {NoStop}%
\bibitem [{\citenamefont {Shin}\ \emph
  {et~al.}(2008{\natexlab{a}})\citenamefont {Shin}, \citenamefont {Schirotzek},
  \citenamefont {Schunck},\ and\ \citenamefont
  {Ketterle}}]{PhysRevLett.101.070404}%
  \BibitemOpen
  \bibfield  {author} {\bibinfo {author} {\bibfnamefont {Y.-i.}\ \bibnamefont
  {Shin}}, \bibinfo {author} {\bibfnamefont {A.}~\bibnamefont {Schirotzek}},
  \bibinfo {author} {\bibfnamefont {C.~H.}\ \bibnamefont {Schunck}},\ and\
  \bibinfo {author} {\bibfnamefont {W.}~\bibnamefont {Ketterle}},\ }\bibfield
  {title} {\bibinfo {title} {Realization of a strongly interacting {Bose-Fermi}
  mixture from a two-component {Fermi} gas},\ }\href
  {https://doi.org/10.1103/PhysRevLett.101.070404} {\bibfield  {journal}
  {\bibinfo  {journal} {Phys. Rev. Lett.}\ }\textbf {\bibinfo {volume} {101}},\
  \bibinfo {pages} {070404} (\bibinfo {year} {2008}{\natexlab{a}})}\BibitemShut
  {NoStop}%
\bibitem [{\citenamefont {Shin}\ \emph
  {et~al.}(2008{\natexlab{b}})\citenamefont {Shin}, \citenamefont {Schunck},
  \citenamefont {Schirotzek},\ and\ \citenamefont {Ketterle}}]{Shin2008}%
  \BibitemOpen
  \bibfield  {author} {\bibinfo {author} {\bibfnamefont {Y.-i.}\ \bibnamefont
  {Shin}}, \bibinfo {author} {\bibfnamefont {C.~H.}\ \bibnamefont {Schunck}},
  \bibinfo {author} {\bibfnamefont {A.}~\bibnamefont {Schirotzek}},\ and\
  \bibinfo {author} {\bibfnamefont {W.}~\bibnamefont {Ketterle}},\ }\bibfield
  {title} {\bibinfo {title} {Phase diagram of a two-component {Fermi} gas with
  resonant interactions},\ }\href {https://doi.org/10.1038/nature06473}
  {\bibfield  {journal} {\bibinfo  {journal} {Nature}\ }\textbf {\bibinfo
  {volume} {451}},\ \bibinfo {pages} {689} (\bibinfo {year}
  {2008}{\natexlab{b}})}\BibitemShut {NoStop}%
\bibitem [{\citenamefont {Mordini}\ \emph {et~al.}(2020)\citenamefont
  {Mordini}, \citenamefont {Trypogeorgos}, \citenamefont {Farolfi},
  \citenamefont {Wolswijk}, \citenamefont {Stringari}, \citenamefont
  {Lamporesi},\ and\ \citenamefont {Ferrari}}]{PhysRevLett.125.150404}%
  \BibitemOpen
  \bibfield  {author} {\bibinfo {author} {\bibfnamefont {C.}~\bibnamefont
  {Mordini}}, \bibinfo {author} {\bibfnamefont {D.}~\bibnamefont
  {Trypogeorgos}}, \bibinfo {author} {\bibfnamefont {A.}~\bibnamefont
  {Farolfi}}, \bibinfo {author} {\bibfnamefont {L.}~\bibnamefont {Wolswijk}},
  \bibinfo {author} {\bibfnamefont {S.}~\bibnamefont {Stringari}}, \bibinfo
  {author} {\bibfnamefont {G.}~\bibnamefont {Lamporesi}},\ and\ \bibinfo
  {author} {\bibfnamefont {G.}~\bibnamefont {Ferrari}},\ }\bibfield  {title}
  {\bibinfo {title} {Measurement of the canonical equation of state of a weakly
  interacting 3d {Bose} gas},\ }\href
  {https://doi.org/10.1103/PhysRevLett.125.150404} {\bibfield  {journal}
  {\bibinfo  {journal} {Phys. Rev. Lett.}\ }\textbf {\bibinfo {volume} {125}},\
  \bibinfo {pages} {150404} (\bibinfo {year} {2020})}\BibitemShut {NoStop}%
\bibitem [{\citenamefont {Navon}\ \emph {et~al.}(2021)\citenamefont {Navon},
  \citenamefont {Smith},\ and\ \citenamefont {Hadzibabic}}]{Navon2021}%
  \BibitemOpen
  \bibfield  {author} {\bibinfo {author} {\bibfnamefont {N.}~\bibnamefont
  {Navon}}, \bibinfo {author} {\bibfnamefont {R.~P.}\ \bibnamefont {Smith}},\
  and\ \bibinfo {author} {\bibfnamefont {Z.}~\bibnamefont {Hadzibabic}},\
  }\bibfield  {title} {\bibinfo {title} {Quantum gases in optical boxes},\
  }\href {https://doi.org/10.1038/s41567-021-01403-z} {\bibfield  {journal}
  {\bibinfo  {journal} {Nat. Phys.}\ }\textbf {\bibinfo {volume} {17}},\
  \bibinfo {pages} {1334} (\bibinfo {year} {2021})}\BibitemShut {NoStop}%
\bibitem [{\citenamefont {He}\ \emph {et~al.}(2023{\natexlab{b}})\citenamefont
  {He}, \citenamefont {Li}, \citenamefont {Yi},\ and\ \citenamefont
  {Yu}}]{PhysRevLett.130.193001}%
  \BibitemOpen
  \bibfield  {author} {\bibinfo {author} {\bibfnamefont {L.}~\bibnamefont
  {He}}, \bibinfo {author} {\bibfnamefont {H.}~\bibnamefont {Li}}, \bibinfo
  {author} {\bibfnamefont {W.}~\bibnamefont {Yi}},\ and\ \bibinfo {author}
  {\bibfnamefont {Z.-Q.}\ \bibnamefont {Yu}},\ }\bibfield  {title} {\bibinfo
  {title} {Quantum criticality of liquid-gas transition in a binary bose
  mixture},\ }\href {https://doi.org/10.1103/PhysRevLett.130.193001} {\bibfield
   {journal} {\bibinfo  {journal} {Phys. Rev. Lett.}\ }\textbf {\bibinfo
  {volume} {130}},\ \bibinfo {pages} {193001} (\bibinfo {year}
  {2023}{\natexlab{b}})}\BibitemShut {NoStop}%
\bibitem [{\citenamefont {Spada}\ \emph {et~al.}(2023)\citenamefont {Spada},
  \citenamefont {Pilati},\ and\ \citenamefont {Giorgini}}]{spada_2023_7845396}%
  \BibitemOpen
  \bibfield  {author} {\bibinfo {author} {\bibfnamefont {G.}~\bibnamefont
  {Spada}}, \bibinfo {author} {\bibfnamefont {S.}~\bibnamefont {Pilati}},\ and\
  \bibinfo {author} {\bibfnamefont {S.}~\bibnamefont {Giorgini}},\ }\bibfield
  {title} {\bibinfo {title} {{Data for: Attractive solution of binary Bose
  mixtures: Liquid-vapor coexistence and critical point}},\ }\href
  {https://doi.org/10.5281/zenodo.7845396} {10.5281/zenodo.7845396} (\bibinfo
  {year} {2023})\BibitemShut {NoStop}%
\bibitem [{\citenamefont {Pitaevskii}\ and\ \citenamefont
  {Stringari}(2016)}]{book}%
  \BibitemOpen
  \bibfield  {author} {\bibinfo {author} {\bibfnamefont {L.}~\bibnamefont
  {Pitaevskii}}\ and\ \bibinfo {author} {\bibfnamefont {S.}~\bibnamefont
  {Stringari}},\ }\href@noop {} {\emph {\bibinfo {title} {{Bose-Einstein}
  Condensation and Superfluidity}}}\ (\bibinfo  {publisher} {Oxford University
  Press, New York},\ \bibinfo {year} {2016})\BibitemShut {NoStop}%
\end{thebibliography}%


\begin{thebibliography}{12}%
\makeatletter
\providecommand \@ifxundefined [1]{%
 \@ifx{#1\undefined}
}%
\providecommand \@ifnum [1]{%
 \ifnum #1\expandafter \@firstoftwo
 \else \expandafter \@secondoftwo
 \fi
}%
\providecommand \@ifx [1]{%
 \ifx #1\expandafter \@firstoftwo
 \else \expandafter \@secondoftwo
 \fi
}%
\providecommand \natexlab [1]{#1}%
\providecommand \enquote  [1]{``#1''}%
\providecommand \bibnamefont  [1]{#1}%
\providecommand \bibfnamefont [1]{#1}%
\providecommand \citenamefont [1]{#1}%
\providecommand \href@noop [0]{\@secondoftwo}%
\providecommand \href [0]{\begingroup \@sanitize@url \@href}%
\providecommand \@href[1]{\@@startlink{#1}\@@href}%
\providecommand \@@href[1]{\endgroup#1\@@endlink}%
\providecommand \@sanitize@url [0]{\catcode `\\12\catcode `\$12\catcode
  `\&12\catcode `\#12\catcode `\^12\catcode `\_12\catcode `\%12\relax}%
\providecommand \@@startlink[1]{}%
\providecommand \@@endlink[0]{}%
\providecommand \url  [0]{\begingroup\@sanitize@url \@url }%
\providecommand \@url [1]{\endgroup\@href {#1}{\urlprefix }}%
\providecommand \urlprefix  [0]{URL }%
\providecommand \Eprint [0]{\href }%
\providecommand \doibase [0]{https://doi.org/}%
\providecommand \selectlanguage [0]{\@gobble}%
\providecommand \bibinfo  [0]{\@secondoftwo}%
\providecommand \bibfield  [0]{\@secondoftwo}%
\providecommand \translation [1]{[#1]}%
\providecommand \BibitemOpen [0]{}%
\providecommand \bibitemStop [0]{}%
\providecommand \bibitemNoStop [0]{.\EOS\space}%
\providecommand \EOS [0]{\spacefactor3000\relax}%
\providecommand \BibitemShut  [1]{\csname bibitem#1\endcsname}%
\let\auto@bib@innerbib\@empty
\bibitem [{\citenamefont {Ceperley}(1995)}]{RevModPhys.67.279}%
  \BibitemOpen
  \bibfield  {author} {\bibinfo {author} {\bibfnamefont {D.~M.}\ \bibnamefont
  {Ceperley}},\ }\bibfield  {title} {\bibinfo {title} {{Path integrals in the
  theory of condensed helium}},\ }\href
  {https://doi.org/10.1103/RevModPhys.67.279} {\bibfield  {journal} {\bibinfo
  {journal} {Rev. Mod. Phys.}\ }\textbf {\bibinfo {volume} {67}},\ \bibinfo
  {pages} {279} (\bibinfo {year} {1995})}\BibitemShut {NoStop}%
\bibitem [{\citenamefont {Pilati}\ \emph {et~al.}(2006)\citenamefont {Pilati},
  \citenamefont {Sakkos}, \citenamefont {Boronat}, \citenamefont {Casulleras},\
  and\ \citenamefont {Giorgini}}]{PhysRevA.74.043621}%
  \BibitemOpen
  \bibfield  {author} {\bibinfo {author} {\bibfnamefont {S.}~\bibnamefont
  {Pilati}}, \bibinfo {author} {\bibfnamefont {K.}~\bibnamefont {Sakkos}},
  \bibinfo {author} {\bibfnamefont {J.}~\bibnamefont {Boronat}}, \bibinfo
  {author} {\bibfnamefont {J.}~\bibnamefont {Casulleras}},\ and\ \bibinfo
  {author} {\bibfnamefont {S.}~\bibnamefont {Giorgini}},\ }\bibfield  {title}
  {\bibinfo {title} {{Equation of state of an interacting Bose gas at finite
  temperature: A path-integral Monte Carlo study}},\ }\href
  {https://doi.org/10.1103/PhysRevA.74.043621} {\bibfield  {journal} {\bibinfo
  {journal} {Phys. Rev. A}\ }\textbf {\bibinfo {volume} {74}},\ \bibinfo
  {pages} {043621} (\bibinfo {year} {2006})}\BibitemShut {NoStop}%
\bibitem [{\citenamefont {Spada}\ \emph
  {et~al.}(2022{\natexlab{a}})\citenamefont {Spada}, \citenamefont {Pilati},\
  and\ \citenamefont {Giorgini}}]{PhysRevA.105.013325}%
  \BibitemOpen
  \bibfield  {author} {\bibinfo {author} {\bibfnamefont {G.}~\bibnamefont
  {Spada}}, \bibinfo {author} {\bibfnamefont {S.}~\bibnamefont {Pilati}},\ and\
  \bibinfo {author} {\bibfnamefont {S.}~\bibnamefont {Giorgini}},\ }\bibfield
  {title} {\bibinfo {title} {Thermodynamics of a dilute {Bose} gas: A
  path-integral {Monte Carlo} study},\ }\href
  {https://doi.org/10.1103/PhysRevA.105.013325} {\bibfield  {journal} {\bibinfo
   {journal} {Phys. Rev. A}\ }\textbf {\bibinfo {volume} {105}},\ \bibinfo
  {pages} {013325} (\bibinfo {year} {2022}{\natexlab{a}})}\BibitemShut
  {NoStop}%
\bibitem [{\citenamefont {Boninsegni}\ \emph {et~al.}(2006)\citenamefont
  {Boninsegni}, \citenamefont {Prokof'ev},\ and\ \citenamefont
  {Svistunov}}]{PhysRevLett.96.070601}%
  \BibitemOpen
  \bibfield  {author} {\bibinfo {author} {\bibfnamefont {M.}~\bibnamefont
  {Boninsegni}}, \bibinfo {author} {\bibfnamefont {N.}~\bibnamefont
  {Prokof'ev}},\ and\ \bibinfo {author} {\bibfnamefont {B.}~\bibnamefont
  {Svistunov}},\ }\bibfield  {title} {\bibinfo {title} {{Worm algorithm for
  continuous-space path integral Monte Carlo simulations}},\ }\href
  {https://doi.org/10.1103/PhysRevLett.96.070601} {\bibfield  {journal}
  {\bibinfo  {journal} {Phys. Rev. Lett.}\ }\textbf {\bibinfo {volume} {96}},\
  \bibinfo {pages} {070601} (\bibinfo {year} {2006})}\BibitemShut {NoStop}%
\bibitem [{\citenamefont {Spada}\ \emph
  {et~al.}(2022{\natexlab{b}})\citenamefont {Spada}, \citenamefont {Giorgini},\
  and\ \citenamefont {Pilati}}]{condmat7020030}%
  \BibitemOpen
  \bibfield  {author} {\bibinfo {author} {\bibfnamefont {G.}~\bibnamefont
  {Spada}}, \bibinfo {author} {\bibfnamefont {S.}~\bibnamefont {Giorgini}},\
  and\ \bibinfo {author} {\bibfnamefont {S.}~\bibnamefont {Pilati}},\
  }\bibfield  {title} {\bibinfo {title} {Path-integral {Monte Carlo} worm
  algorithm for {Bose} systems with periodic boundary conditions},\ }\href
  {https://doi.org/10.3390/condmat7020030} {\bibfield  {journal} {\bibinfo
  {journal} {Condens. Matter}\ }\textbf {\bibinfo {volume} {7}},\ \bibinfo
  {pages} {30} (\bibinfo {year} {2022}{\natexlab{b}})}\BibitemShut {NoStop}%
\bibitem [{\citenamefont {Cao}\ and\ \citenamefont
  {Berne}(1992)}]{doi:10.1063/1.463076}%
  \BibitemOpen
  \bibfield  {author} {\bibinfo {author} {\bibfnamefont {J.}~\bibnamefont
  {Cao}}\ and\ \bibinfo {author} {\bibfnamefont {B.~J.}\ \bibnamefont
  {Berne}},\ }\bibfield  {title} {\bibinfo {title} {{A new quantum propagator
  for hard sphere and cavity systems}},\ }\href
  {https://doi.org/10.1063/1.463076} {\bibfield  {journal} {\bibinfo  {journal}
  {J. Chem. Phys.}\ }\textbf {\bibinfo {volume} {97}},\ \bibinfo {pages} {2382}
  (\bibinfo {year} {1992})}\BibitemShut {NoStop}%
\bibitem [{\citenamefont {Allen}\ and\ \citenamefont
  {Tildesley}(1987)}]{AllenTildesley}%
  \BibitemOpen
  \bibfield  {author} {\bibinfo {author} {\bibfnamefont {M.~P.}\ \bibnamefont
  {Allen}}\ and\ \bibinfo {author} {\bibfnamefont {D.~J.}\ \bibnamefont
  {Tildesley}},\ }\href@noop {} {\emph {\bibinfo {title} {Computer Simulations
  of Liquids}}}\ (\bibinfo  {publisher} {Oxford University Press, New York},\
  \bibinfo {year} {1987})\BibitemShut {NoStop}%
\bibitem [{\citenamefont {Alder}\ and\ \citenamefont
  {Wainwright}(1962)}]{Alder1962dis}%
  \BibitemOpen
  \bibfield  {author} {\bibinfo {author} {\bibfnamefont {B.~J.}\ \bibnamefont
  {Alder}}\ and\ \bibinfo {author} {\bibfnamefont {T.~E.}\ \bibnamefont
  {Wainwright}},\ }\bibfield  {title} {\bibinfo {title} {Phase transition in
  elastic disks},\ }\href {https://doi.org/10.1103/PhysRev.127.359} {\bibfield
  {journal} {\bibinfo  {journal} {Phys. Rev.}\ }\textbf {\bibinfo {volume}
  {127}},\ \bibinfo {pages} {359} (\bibinfo {year} {1962})}\BibitemShut
  {NoStop}%
\bibitem [{\citenamefont {Reif-Acherman}(2010)}]{reif2010history}%
  \BibitemOpen
  \bibfield  {author} {\bibinfo {author} {\bibfnamefont {S.}~\bibnamefont
  {Reif-Acherman}},\ }\bibfield  {title} {\bibinfo {title} {The history of the
  rectilinear diameter law},\ }\href@noop {} {\bibfield  {journal} {\bibinfo
  {journal} {Quim. Nova}\ }\textbf {\bibinfo {volume} {33}},\ \bibinfo {pages}
  {2003} (\bibinfo {year} {2010})}\BibitemShut {NoStop}%
\bibitem [{\citenamefont {Cornfeld}\ and\ \citenamefont
  {Carr}(1972)}]{PhysRevLett.29.28}%
  \BibitemOpen
  \bibfield  {author} {\bibinfo {author} {\bibfnamefont {A.~B.}\ \bibnamefont
  {Cornfeld}}\ and\ \bibinfo {author} {\bibfnamefont {H.~Y.}\ \bibnamefont
  {Carr}},\ }\bibfield  {title} {\bibinfo {title} {Experimental evidence
  concerning the law of rectilinear diameter},\ }\href
  {https://doi.org/10.1103/PhysRevLett.29.28} {\bibfield  {journal} {\bibinfo
  {journal} {Phys. Rev. Lett.}\ }\textbf {\bibinfo {volume} {29}},\ \bibinfo
  {pages} {28} (\bibinfo {year} {1972})}\BibitemShut {NoStop}%
\bibitem [{\citenamefont {Watanabe}\ \emph {et~al.}(2012)\citenamefont
  {Watanabe}, \citenamefont {Ito},\ and\ \citenamefont
  {Hu}}]{10.1063/1.4720089}%
  \BibitemOpen
  \bibfield  {author} {\bibinfo {author} {\bibfnamefont {H.}~\bibnamefont
  {Watanabe}}, \bibinfo {author} {\bibfnamefont {N.}~\bibnamefont {Ito}},\ and\
  \bibinfo {author} {\bibfnamefont {C.-K.}\ \bibnamefont {Hu}},\ }\bibfield
  {title} {\bibinfo {title} {{Phase diagram and universality of the
  {Lennard-Jones} gas-liquid system}},\ }\bibfield  {journal} {\bibinfo
  {journal} {J. Chem. Phys.}\ }\textbf {\bibinfo {volume} {136}},\ \href
  {https://doi.org/10.1063/1.4720089} {10.1063/1.4720089} (\bibinfo {year}
  {2012}),\ \bibinfo {note} {204102}\BibitemShut {NoStop}%
\bibitem [{\citenamefont {Rovere}\ \emph {et~al.}(1990)\citenamefont {Rovere},
  \citenamefont {Heermann},\ and\ \citenamefont {Binder}}]{Rovere_1990}%
  \BibitemOpen
  \bibfield  {author} {\bibinfo {author} {\bibfnamefont {M.}~\bibnamefont
  {Rovere}}, \bibinfo {author} {\bibfnamefont {D.~W.}\ \bibnamefont
  {Heermann}},\ and\ \bibinfo {author} {\bibfnamefont {K.}~\bibnamefont
  {Binder}},\ }\bibfield  {title} {\bibinfo {title} {The gas-liquid transition
  of the two-dimensional {Lennard-Jones} fluid},\ }\href
  {https://doi.org/10.1088/0953-8984/2/33/013} {\bibfield  {journal} {\bibinfo
  {journal} {Journal of Physics: Condensed Matter}\ }\textbf {\bibinfo {volume}
  {2}},\ \bibinfo {pages} {7009} (\bibinfo {year} {1990})}\BibitemShut
  {NoStop}%
\end{thebibliography}%

\end{document}


\author{G. Spada$^1$, S. Pilati$^{2,3}$, and S. Giorgini$^1$\vspace*{10pt}}

\affiliation{
    $^1$ Pitaevskii Center on Bose-Einstein Condensation, CNR-INO and Dipartimento di Fisica, Universit\`a di Trento, 38123 Povo, Trento, Italy \\
    $^2$ School of Science and Technology, Physics Division, Universit\`a di Camerino, 62032 Camerino, Italy \\
    $^3$ INFN, Sezione di Perugia, I-06123 Perugia, Italy
}

\title{Supplemental Material for\\``Attractive Solution of Binary Bose Mixtures: \\
Liquid-Vapour Coexistence and Critical Point ''}

\maketitle

\setcounter{figure}{0}
\setcounter{equation}{0}
\renewcommand\thefigure{S\arabic{figure}}
\renewcommand\theequation{S\arabic{equation}}
\renewcommand{\bibnumfmt}[1]{[S#1]}
\renewcommand{\citenumfont}[1]{S#1}

\section{PIMC for binary mixtures}

In a PIMC simulation one samples multidimensional configurations ${\bf R}=(\dots{\bf r}_i\dots,\dots{\bf r}_{i^\prime}\dots)$, where $i$ ($i^\prime$) runs over the $N_1$ ($N_2$) particles of each component, distributed according to the density matrix $\rho({\bf R},P{\bf R},\beta)=\langle{\bf R}|e^{-\beta H}|P{\bf R}\rangle$ at the inverse temperature $\beta=1/k_BT$ and for a generic permutation $P$ of the indistinguishable particles in the two components. The distribution of configurations ${\bf R}$ is normalized to the partition function
\begin{equation}
	Z=\frac{1}{N_1!N_2!}\sum_P\int d{\bf R}\;\rho({\bf R},P{\bf R},\beta)
	\label{partitionfunction}
\end{equation}
and physical observables, such as the pressure $p=\frac{k_BT}{Z}\frac{dZ}{dV}$, are calculated from averages over the probability distribution $\frac{1}{ZN_1!N_2!}\sum_P\rho({\bf R},P{\bf R},\beta)$. By means of the convolution integral the density matrix at temperature $1/\beta$ can be decomposed into $M\gg1$ imaginary time steps, resulting in a product of density matrices at the much higher effective temperature $1/\delta=M/\beta$. Convenient approximations can be used for the density matrix $\rho({\bf R},{\bf R}^\prime,\delta)$~\cite{RevModPhys.67.279}.

In particular, we employ the pair-product approximation~\cite{PhysRevA.74.043621,PhysRevA.105.013325}, based on the solution of the two-body problem for the hard-sphere potential $v(r)$ and the pseudopotential $v_{12}(r)$ (see below for the definition), to sample the partition function $Z$ and compute physical observables, such as the pressure $p=\frac{k_BT}{Z}\frac{\mathrm{d}Z}{\mathrm{d}V}$.
Furthermore, an efficient sampling of the space of particle permutations is provided by the worm algorithm~\cite{PhysRevLett.96.070601}, of which we use an implementation that is particularly well suited to deal with systems subject to periodic boundary conditions~\cite{condmat7020030}.
Using the worm algorithm has the additional benefit of allowing the computation of the one-body density matrix~\cite{PhysRevLett.96.070601}, from which one can extract the condensate density $n_0$ as its limiting value at large distances.
In all the simulations we check the dependence of the results on the number $M$ of imaginary time steps, ensuring the convergence to the $M\to\infty$ limit.

\subsection{Pair product approximation for binary mixtures}

In the pair product approximation the density matrix of the many-body system is decomposed into the contribution from all possible pairs of particles. In the case of a mixture of $N_1$ and $N_2$ particles of the two components the density matrix is written as the product of three terms
\begin{equation}
\rho({\bf R},{\bf R}^\prime,\tau)=\rho_{11}({\bf R},{\bf R}^\prime,\tau)\rho_{22}({\bf R},{\bf R}^\prime,\tau)\rho_{12}({\bf R},{\bf R}^\prime,\tau) \;.
\label{PP1}
\end{equation}
The first two terms refer to pairs of particles of the same component~\cite{RevModPhys.67.279,PhysRevA.74.043621,PhysRevA.105.013325}
\begin{equation}
\rho_{11}({\bf R},{\bf R}^\prime,\tau)=\prod_{i=1}^{N_1}\rho_1({\bf r}_i,{\bf r}_
i^\prime,\tau)\prod_{i<j}
\frac{\rho_{rel}^{\text{HS}}({\bf r}_{ij},{\bf r}_{ij}^\prime,\tau)}
{\rho_{rel}^0({\bf r}_{ij},{\bf r}_{ij}^\prime,\tau)} \;,
\end{equation}
and similarly for $\rho_{22}({\bf R},{\bf R}^\prime,\tau)$. Here $\rho_1$ is the single-particle non-interacting density matrix
\begin{equation}
\rho_1({\bf r}_i,{\bf r}_i^\prime,\tau)=\left(\frac{m}{2\pi\hbar^2\tau}
\right)^{3/2}
e^{-({\bf r}_i-{\bf r}_i^\prime)^2m/(2\hbar^2\tau)} \;,
\label{PP3}
\end{equation}
and $\rho_{rel}^{\text{HS}}$ is the two-body density matrix of the interacting system, which depends on
the relative coordinates  ${\bf r}_{ij}={\bf r}_i-{\bf r}_j$ and ${\bf r}_{ij}^\prime={\bf r}_i^\prime-{\bf r}_j^\prime$,
divided by the corresponding non-interacting term
\begin{equation}
\rho_{rel}^0({\bf r}_{ij},{\bf r}_{ij}^\prime,\tau)=
\left(\frac{m}{4\pi\hbar^2\tau}\right)^{3/2}
e^{-({\bf r}_{ij}-{\bf r}_{ij}^\prime)^2 m/(4\hbar^2\tau)} \;.
\label{PP4}
\end{equation}
Intra-species interactions are modeled by a hard-sphere (HS) potential, where the value $a$ of the range coincides with the corresponding $s$-wave
scattering length. For this potential a simple analytical approximation of the high-temperature two-body density matrix due to Cao and Berne~\cite{doi:10.1063/1.463076}
has been proven to be highly accurate. The approximation is obtained from the large momentum expansion of the solution to the two-body Schr\"odinger
scattering equation yielding the result
\begin{equation}
\frac{\rho_{rel}^{\text{HS}}({\bf r},{\bf r}^\prime,\tau)}
{\rho_{rel}^0({\bf r},{\bf r}^\prime,\tau)}=
 1 -\frac{a(r+r^\prime)-a^2}{rr^\prime} e^{-[rr^\prime +a^2-a(r+r^\prime)](1+\cos\theta)m/(2\hbar^2\tau)} \;.
\label{PP5}
\end{equation}
The contribution to Eq.~\ref{PP1} from the pairs of particles belonging to different components is written as
\begin{equation}
\rho_{12}({\bf R},{\bf R}^\prime,\tau)=\prod_{i=1}^{N_1}\rho_1({\bf r}_i,{\bf r}_
i^\prime,\tau)\prod_{i^\prime=1}^{N_2}\rho_1({\bf r}_{i^\prime},{\bf r}_{i^\prime}^\prime,\tau)
\prod_{i,i^\prime}
\frac{\rho_{rel}^{\text{PP}}({\bf r}_{ii^\prime},{\bf r}_{ii^\prime}^\prime,\tau)}
{\rho_{rel}^0({\bf r}_{ii^\prime},{\bf r}_{ii^\prime}^\prime,\tau)} \;.
\end{equation}
Such pairs of particles interact via the potential $v_{12}({\bf r})$ for which we use a pseudopotential model $v_{12}({\bf r})=\frac{4\pi\hbar^2a_{12}}{m}\delta({\bf r}) \frac{\partial}{\partial r}\left(r\cdot \; \right)$
characterized by the negative scattering length $a_{12}<0$. The solution of the two-body scattering problem involves in this case only the $s$-wave channel, and the corresponding density matrix can be evaluated in closed form
\begin{equation}
    \frac{\rho_{rel}^{\mathrm{PP}}({\bf r},{\bf r}^\prime,\tau)}
    {\rho_{rel}^0({\bf r},{\bf r}^\prime,\tau)}
    =
    1 + \frac{2 \hbar^2 \tau}{m r r^\prime}
    e^{-\frac{m}{2 \hbar^2 \tau} r r^\prime (1+\cos\theta)}
    \left[
        1 + \frac{1}{a_{12}} \sqrt{\frac{\pi \hbar^2 \tau}{m} } e^{z^2} \erfc(z)
    \right]\,,
    \label{eq:rhoPP}
\end{equation}
where $\erfc(z)=\frac{2}{\sqrt{\pi}}\int_z^\infty dt\;e^{-t^2}$ is the complementary error function and we have defined
\begin{equation}
    z = \sqrt{\frac{\tau\hbar^2}{ma_{12}^2}}
    + \frac{1}{2} \frac{r+r^\prime}{\sqrt{\tau\hbar^2/m}}
    \,.
\end{equation}
Notice that Eq.~\eqref{eq:rhoPP} involves the product $e^{z^2} \erfc(z)$ that can lead to loss of precision when evaluated numerically. Indeed, for moderate values of $z$, the exponential overflows, while the complementary error function underflows. To avoid the problem one can use a numerically stable implementation of the scaled complementary error function $\erfcx(z) = e^{z^2} \erfc(z)$.

One should notice that, according to our choice in Eq.~\eqref{eq:rhoPP}, interspecies interactions are short range and purely attractive. As a consequence, the stabilizing mechanism avoiding collapse of the mixture and leading to the gas to liquid transition is provided by repulsive beyond mean-field effects within each component. Our results show that a similar balance between attractive mean-field and repulsive beyond mean-field interaction terms occurring for the ground state is also active at finite temperature.

\begin{figure}[t]
\centering
    \includegraphics[width=0.55\textwidth]{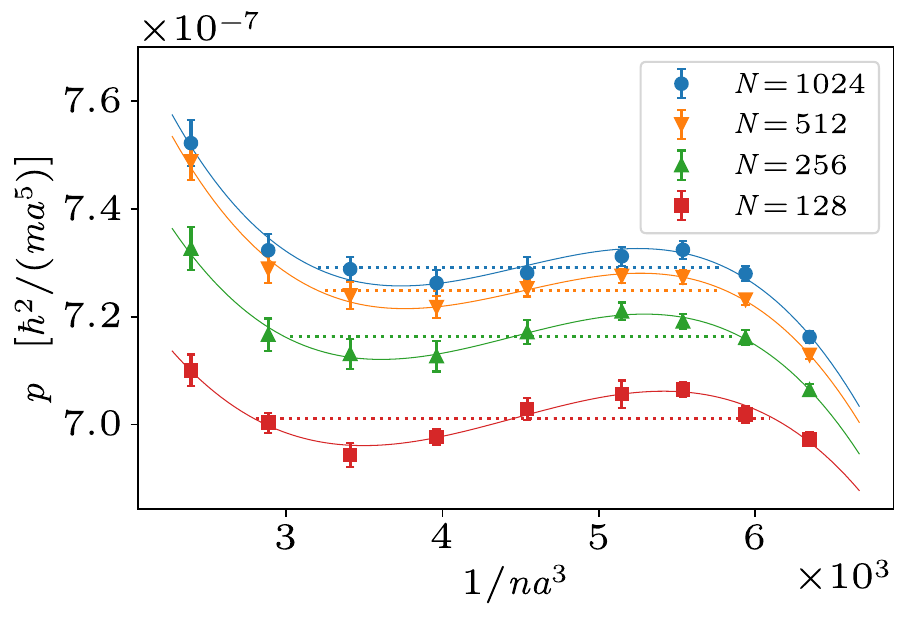}
    \caption{Pressure for finite-size systems in the coexistence region at the temperature $k_B T = 0.007 \, \hbar /(ma^2)$ and interspecies interaction $a_{12}=-1.1a$. The PIMC points for four system sizes have been interpolated with cubic polynomials (solid lines). The horizontal dotted lines show the values of the pressure and the limiting densities of the coexistence region as obtained with the Maxwell construction (central values).}
    \label{figS1}
\end{figure}

\section{Finite size effects and Maxwell construction}

Near the first-order transition in finite systems, the pressure-volume plot exhibits a characteristic S-shaped behavior for the isothermal lines, (see e.g.~\cite{AllenTildesley}). However, determining the coexistence region and vapor pressure in the thermodynamic limit is challenging due to the difficulty in reliably extrapolating the data. To overcome this issue, we adopt the Maxwell construction method on finite-size systems to exploit the non-monotonicity of the curves. By fitting the PIMC data to a cubic function, we can accurately determine the coexistence region and vapor pressure, as detailed in the main text.
Figure~\ref{figS1} shows the pressure data for four different system sizes, along with the corresponding cubic fits and the vapor pressure, at $k_B T = 0.007 \, \hbar^2/(ma^2)$ for a mixture with $a_{12}=-1.1a$. We observe that the S-shaped behavior becomes less pronounced and more flattened as the system size increases from $N=128$ to $N=1024$. Furthermore, the coexistence region obtained from $N=1024$ is quantitatively consistent with that obtained from $N=512$, indicating that $N=512$ is a sufficient system size to yield results close to the thermodynamic limit.\\

It is worth pointing out that, in principle, simulating sufficiently large boxes should allow one to discern the two coexisting phases in different regions of the box. For example, the coexistence between liquid and crystal phases was identified in Ref.~\cite{Alder1962dis}, in the case of classical hard disks, by observing snapshots of particle trajectories. However, for the system sizes we address, it is not feasible to discern liquid and gas regions, since both feature disordered structures, as opposed to liquid and crystal phases.

\section{Critical point determination}

\begin{figure}
\centering
    \includegraphics[width=0.55\textwidth]{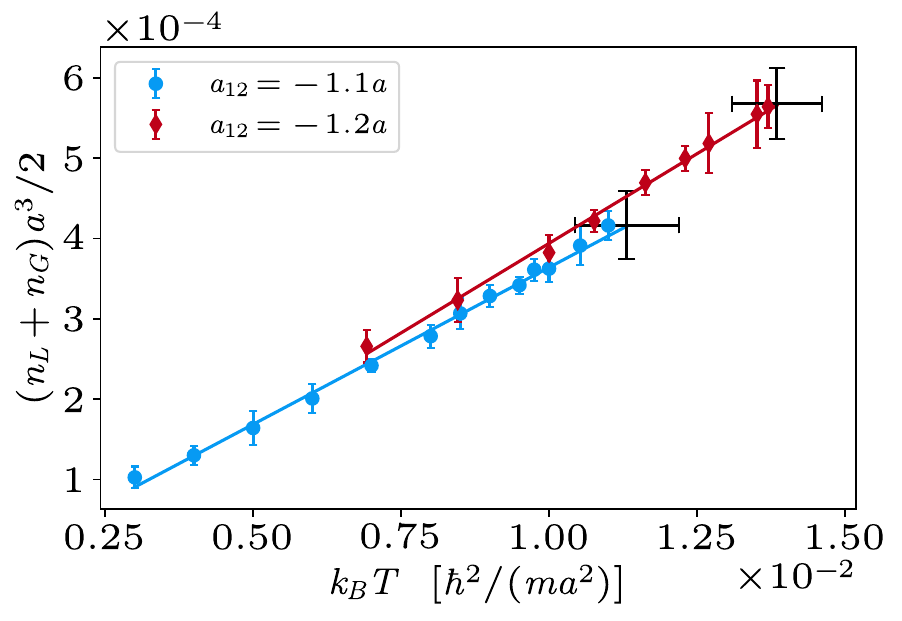}
    \caption{Linear extrapolation of the mean density in the coexistence region, to extract the density of the critical points (black error bar points).}
    \label{figS2}
\end{figure}

The densities for the liquid and gas phases (denoted respectively $n_L$ and $n_G$), which bound the coexistence region, have been determined exploiting the non-monotonic behavior of the isotherms. Above the temperature of the critical point, the first-order transition line terminates and the isotherms become monotonic, allowing us to pinpoint the temperature $T_c$ of the critical point (within a confidence interval). In order to extract the critical density, we linearly extrapolate the mean density of the coexistence region as a function of the temperature up to our estimate of $T_c$.
In fact, this linear dependence is observed in several examples of classical liquid-gas coexistence. It is referred to as the law of rectilinear diameters~\cite{reif2010history,PhysRevLett.29.28,10.1063/1.4720089}. It is interesting to recover it here in the case of a quantum mixture.
In Fig.~\ref{figS2} we show the extrapolations for the two values of $a_{12}$ and the resulting critical points, represented by the black error bars. In order to get the pressure $p_c$ at the critical point, we use a similar procedure, where we extrapolate the vapor pressure to $T_c$.
The critical behavior could be further analyzed using the fourth-order cumulant of the block density distribution, as in previous studies on the classical liquid-gas coexistence~\cite{Rovere_1990}.

\bibliography{CriticalPoint}